\documentclass[article,structabstract]{aa}
\usepackage{graphicx}

\usepackage{amsmath}
\usepackage{txfonts}

\usepackage{multirow}
\usepackage{xspace}
\usepackage{verbatim} 
\usepackage{natbib}
\bibpunct{(}{)}{;}{a}{}{,} 
\newcommand{\um}{\ensuremath{\rm \mu m}\xspace}
\newcommand{\e}[1]{\ensuremath{\rm ~\times~10^{#1}}\xspace}
\newcommand{\ten}[1]{\ensuremath{\rm 10^{#1}}\xspace}
\newcommand{\fit}{\ensuremath{^{\dagger}}\xspace}
\newcommand{\ong}{\ensuremath{\sim}\xspace}
\begin{document}
\title{Low abundance, strong features: Window-dressing crystalline forsterite in the disk wall of HD 100546}
\titlerunning{forsterite in the disk of HD 100546}
 \author{G.D.~Mulders\inst{1,2}
\and L.B.F.M.~Waters\inst{1,3}
\and C.~Dominik\inst{1,5}
\and B.~Sturm\inst{6}
\and J.~Bouwman\inst{6}
\and M.~Min\inst{7}
\and A.P.~Verhoeff\inst{1}
\and B.~Acke \inst{4}\fnmsep\thanks{Postdoctoral Fellow of the Fund for Scientific Research, Flanders}
\and J.~C.~Augereau\inst{8}
\and N.J.~Evans~II\inst{9}
\and Th.~Henning\inst{6}
\and G.~Meeus \inst{10}
\and J.~Olofsson\inst{6}
}

\institute{
Astronomical Institute ``Anton Pannekoek'', University of Amsterdam,
 PO Box 94249, 1090 GE Amsterdam, The Netherlands
\and 
SRON Netherlands Institute for Space Research, PO Box 800, 9700 AV,
Groningen, The Netherlands
\and 
SRON Netherlands Institute for Space Research, Sorbonnelaan 2, 3584 CA Utrecht, The Netherlands
\and 
Instituut voor Sterrenkunde, K.U.Leuven, Celestijnenlaan 200D, B-3001
Leuven, Belgium
\and 
Department of Astrophysics/IMAPP, Radboud University Nijmegen,
P.O. Box 9010 6500 GL Nijmegen The Netherlands
\and 
Max Planck Institute for Astronomy, K\"onigstuhl 17, D-69117
 Heidelberg, Germany
\and 
Astronomical Institute Utrecht, University of Utrecht, PO Box 80000, 3508 TA Utrecht, The Netherlands
\and 
UJF-Grenoble 1 / CNRS-INSU, Institut de Plan\'etologie et d'Astrophysique de Grenoble (IPAG) UMR 5274, Grenoble, F-38041, France
\and 
The University of Texas at Austin, Department of Astronomy,
 1 University Station C1400, Austin, Texas 78712-0259, USA
\and 
Dpt. de F\'{i}sica Te\'{o}rica, Fac. de Ciencias, Universidad
 Aut\'{o}noma de Madrid, Cantoblanco, 28049 Madrid, Spain
} 

  \date{Received 23-02-2011 / Accepted 19-05-2011 }
    \offprints{G.D.Mulders, \email{\bf{mulders@uva.nl}}}

    \abstract{
      Forsterite is one of the crystalline dust species that is often observed in protoplanetary disks and solar system comets. Being absent in the interstellar medium, it must be produced during the disk lifetime. It can therefore serve as a tracer of dust processing and disk evolution, which can lead to a better understanding of the physical processes occurring in the disk, and possibly planet formation. However, the connection of these processes with the overall disk crystallinity remains unclear.
    }{
      We aim to characterize the forsterite abundance and spatial distribution in the disk of the Herbig Be star \object{HD 100546}, to investigate if a connection exists with the large disk gap.
    }{
      We use a 2D radiative transfer code, MCMax, to model the circumstellar dust around HD 100546. We use VISIR Q-band imaging to probe the outer disk geometry and mid-infrared features to model the spatial distribution of forsterite. The temperature-dependent shape of the 69 \um feature observed with Herschel\thanks{{\it Herschel} is an ESA space observatory with science instruments provided by European-led Principal Investigator consortia and with important participation from NASA.} PACS is used as a critical tool to constrain this distribution.
    }{
      We find a crystalline mass fraction of 40 - 60 \%, located close to the disk wall between 13 and 20 AU, and possibly farther out at the disk surface. The forsterite is in thermal contact with the other dust species. We put an upper limit on the iron content of forsterite of 0.3 \%.
    }{
      Optical depth effects play a key role in explaining the observed forsterite features, hiding warm forsterite from view at short wavelengths. The disk wall acts as a showcase: it displays a localized high abundance of forsterite, which gives rise to a high observed crystallinity, while the overall mass fraction of forsterite is a factor of ten lower.
      }

      \maketitle

      \section{Introduction}
      Herbig Ae/Be stars are intermediate-mass pre-main-sequence stars first described as a group by Herbig (1960). They are characterized by the presence of a circumstellar gas and dust disk \citep[e.g.][]{1998ARA&A..36..233W}, which is the remnant of the star formation process. There is by now a wealth of observational evidence suggesting that these disks are the site of planet formation. For instance, imaging at optical, infrared and millimeter wavelengths shows the presence of disk gaps and/or inner holes \citep[e.g. ][]{2005ApJ...620..470G, 2011arXiv1101.5719V}, pointing to clearing of substantial parts of the disk. The composition of the dust in the disk  strongly differs from that of dust in the interstellar medium. Large, millimeter sized dust grains have grown in the disk and settled to the mid-plane, depleting the disk of small, sub-micron sized particles. In addition, crystalline silicates are detected in the mid-infrared spectra of many disks, pointing to substantial grain processing \citep[e.g. ][]{2001A&A...375..950B, 2005A&A...437..189V}. The recent discovery by direct imaging of exo-planets orbiting intermediate-mass stars (e.g. \citealt{2008Sci...322.1348M, 2008Sci...322.1345K, 2010Sci...329...57L}) provides strong support for the interpretation that disks surrounding Herbig Ae/Be stars are in the process of planet formation.

      However, many basic questions related to planet formation need still to be clarified. For example: what is the main channel for planet formation as a function of stellar mass? How does planet formation affect the composition of the gas and dust and its spatial distribution? Which observed properties of disks can be used as a signpost of planet formation, and can we derive an empirical evolutionary sequence towards mature planetary systems? In order to answer these questions, multi-wavelength observations of the spatial structure and of the gas and dust composition of proto-planetary disks are needed. In this study, we focus on the spatial distribution of crystalline silicates in the Herbig Be star \object{HD 100546}. This star has been observed with Herschel PACS as part of the DIGIT\footnote{Dust, Ice and Gas In Time} open time key programme that aims to address these questions \citep{2010A&A...518L.129S}.

      Crystalline silicates are not found in interstellar space \citep{2004ApJ...609..826K} but are abundant in a significant fraction of proto-planetary disks (e.g., \citealt{2009A&A...507..327O,2010ApJ...721..431J}). They must therefore have been formed in situ, and trace the thermal and chemical history of the grains in the disk. They can form by thermal annealing of amorphous silicates \citep{2000A&A...364..282F}, or by direct condensation from the gas phase. Both processes require high temperatures, respectively above the glass temperature ($\sim$1000 K) or near the dust evaporation temperature ($\sim$1500 K). These temperatures are typically found close to the star.

      Crystals are also observed much farther away from the star than expected based on their temperature (e.g., \citealt{2010A&A...520A..39O}) and are abundant in solar system comets, requiring efficient radial mixing or local production in shocks \citep{2002ApJ...565L.109H}, collisions \citep{2010Icar..207...45M}, parent-body processing \citep{2001M&PS...36..975H} or stellar outbursts \citep{2009Natur.459..224A}. Additionally, crystals can also be destroyed by stellar winds \citep{2009A&A...508..247G}. However, the correlation of crystallinity with other stellar parameters such as age and luminosity in a large sample remains unclear (e.g. \citealt{2009ASPC..414...77W,2010ApJ...714..778O,2010ApJ...721..431J}). Detailed studies of individual disks are therefore needed in which spatially resolved information and spectroscopy are combined in order to establish the spatial distribution, abundance and chemical composition of crystalline silicates in relation to the overall disk geometry.

      In this paper we focus on HD 100546, a Herbig Be star of spectral type B9.5Vne with one of the strongest observed crystalline dust features \citep{1998A&A...332L..25M}. Apart from the unusual mineralogy, the Spectral Energy Distribution (SED) of HD 100546 is characterized by a large mid-infrared excess and a modest near-infrared excess. This SED is explained by a disk gap, which creates a frontally illuminated wall at the far edge of the gap with a temperature of \ong 200 K \citep{2003A&A...401..577B}. The existence of the gap has been confirmed by spatially revolved spectroscopy of the gas as well, placing the wall at 13 AU (e.g. \citealt{2005ApJ...620..470G}). The small near-infrared excess is caused by a tenuous inner disk \citep{2010A&A...511A..75B}.

      The mid-infrared spectrum of HD 100546 has been observed with ISO \citep{1998A&A...332L..25M}, TIMMI2 \citep{2005A&A...437..189V} and Spitzer \citep{2010ApJ...721..431J}. The ISO observations revealed very pronounced crystalline silicate features, similar to those of comet \object{Hale-Bopp} \citep{1997Sci...275.1904C}. Comparison to laboratory spectra indicated that the features between 11 and 30 \um can be described by a dust population with a single temperature of 210 K \citep{1998A&A...332L..25M}, coinciding with the disk wall. The longer wavelength features (33 and 69 \um) require an additional cold component ($\sim$50-70 K) to explain the feature strength.

      The ISO spectrum was analysed in more depth by \cite{2003A&A...401..577B}, who study the forsterite abundance and spatial distribution with a radiative transfer model that assumes the dust is optically thin. They find a crystalline mass fraction that increases from 2\% in the inner disk to 19\% in the outer disk - which is inconsistent with radial mixing models that predict a decreasing fraction with radius.

      The TIMMI2 spectrum was analysed using a similar optically thin model, but without taking into account the geometry, by \cite{2005A&A...437..189V}, who find a forsterite mass fraction of 5.0\% in the 10 \um range. The forsterite features have been observed more recently with Spitzer by \cite{2010ApJ...721..431J}. They find a forsterite mass fraction of 5.6\% from the 10 \um spectral range and 5.1\% from the 20 \um spectral range. 

      In addition to the strength, the shape and central wavelength of the features also contain information on the temperature of the emitting dust, as these are temperature-dependent \citep{2006A&A...449..583K}. This is especially true for the 69 \um feature where the feature broadens and shifts to longer wavelengths at higher temperatures (\citealt{2002MNRAS.331L...1B, 2006MNRAS.370.1599S}, Fig. \ref{fig:topac}).

      The 69 \um feature of HD 100546 has been observed in great detail with Herschel PACS as part of the DIGIT open time key programme, and analysed by \cite{2010A&A...518L.129S} using an optically thin approach. The observed central wavelength of the feature is 69.2 \um, whereas cold (50 K), iron-free forsterite\footnotemark{} has a peak at 69.0 \um. \citet{2010A&A...518L.129S} propose two different scenarios for the required wavelength shift of 0.2 \um:

      \footnotetext{Crystalline silicates with olivine stoichiometry are called forsterite in the case they contain only magnesium (Mg$_2$SiO$_4$), and fayalite if they contain only iron (Fe$_2$SiO$_4$). In the case they contain both iron and magnesium (Fe$_x$Mg$_{2-x}$SiO$_4$), they are called olivine. Because of the extremely low iron content of the crystalline olivine (x $\leqslant$ 0.02) used in this paper, and to avoid confusion with amorphous olivines, we will refer to them as forsterite as well. We will use the denomination iron-free / iron-containing when it is necessary to discriminate between both species.}

      In the first scenario, warm iron-free forsterite creates the observed shape. Using a weighted sum of the opacities, they find that pure forsterite with a temperature of 200-150 K has a peak at 69.2 \um and matches the feature shape. This scenario is consistent with the inferred temperature of the disk wall, but strongly overpredicts the features at shorter wavelengths. It is therefore inconsistent with the ISO and Spitzer data, unless optical depth effects play a significant role.

      In the second scenario, \cite{2010A&A...518L.129S} proposed an alternative explanation: colder forsterite, with a temperature around 50-70 K creates the 69 \um feature, and an admixture of a few percent iron shifts the peak longwards to 69.2 \um. Such a model does not overpredict the short wavelength features, consistent with the earliest model from \cite{1998A&A...332L..25M}. Optical constants for forsterite with a few percent iron have not (yet) been measured in the lab, and a compositional fit could not be made. The inferred wavelength shift follows from interpolation of measurements with 0\% and 10\% iron content (see \citealt{2010A&A...518L.129S}).

      The two scenarios lead not only to a different chemical composition of the forsterite, but also to a different location of emission within the disk. Iron-free forsterite originates from the disk wall, whereas iron-containing forsterite comes from colder material farther out - possibly near the disk midplane. Because forsterite has a large number of features over a broad wavelength range, regions with different optical depths are probed at different wavelengths, providing extra diagnostic power. To shed more light on the chemical composition and spatial distribution of forsterite in the disk, we need to explore how feature strengths and shapes over the entire wavelength range are influenced by these optical depth effects.

      To do this, we need to use a 2D radiative transfer code, described in Section \ref{sec:RT_and_dust}, to construct a physical and mineralogical model of the dust surrounding HD 100546. We construct a structural hydrostatic model in section \ref{sec:struct}, which is consistent with a new VISIR mid-infrared image (appendix \ref{sec:visir}). In section \ref{sec:fors} we use this structural model to study the spatial distribution of crystalline forsterite, by fitting the forsterite feature strengths over a broad wavelength range and in particular the shape of the 69 \um feature.

      We will discuss our results in the context of previous work in section \ref{sec:discussion}, as well as address the question of the origin of the forsterite and correlation with the disk gap. We will summarize our conclusions in section \ref{sec:conclusion}.

      \section{Observations}\label{sec:obs}

      \subsection{SED and spectroscopy}
      To construct the observed SED of \object{HD100546} (see Figure \ref{fig:sed}), we use photometric data from \cite{1997A&A...324L..33V} combined with the ISO spectrum from \cite{1998A&A...332L..25M}. We do not use the Spitzer \citep{2010ApJ...721..431J} and PACS \citep{2010A&A...518L.129S} spectra for constructing the SED because the ISO spectra encompasses the wavelength range of both, and the current flux calibration of the PACS spectrum is not yet better than that of ISO. There is no significant offset between the ISO and Spitzer spectra that could influence our analysis. We do make use of the increased sensitivity of PACS and Spitzer for calculating the integrated strength of the forsterite features.

      \section{Radiative transfer and dust model}\label{sec:RT_and_dust}
      \subsection{Radiative transfer model}\label{sec:RT}
      The dust radiative transfer code used in this paper is MCMax \citep{2009A&A...497..155M}, a 2D Monte Carlo code. It is based on the immediate re-emission procedure from \cite{2001ApJ...554..615B}, combined with the method of continuous absorption by \cite{1999A&A...345..211L}. The code has been benchmarked against other radiative transfer codes for modelling protoplanetary disks \citep{2009A&A...498..967P} and has been successfully applied for modelling spatially and spectrally resolved observations \citep[e.g. ][]{2010A&A...516A..48V}. The radial grid around the inner radius and disk wall is refined to sample the optical depth logarithmically. The SED, images and flux contributions (see section \ref{sec:location}) are calculated by integrating the formal solution to the equation of radiative transfer by ray-tracing.

      In addition to radiative transfer, the code explicitly solves for the hydrostatic vertical structure of the disk, with the implicit assumption that the gas temperature is set by the dust temperature\footnote{The gas scale height is therefore not a free parameter in our model.}. Therefore it only needs the radial distribution of dust to construct the structural model. For the radiative transfer that sets the dust temperature - and therefore its vertical structure - also the stellar and dust properties are required.

      The radial distribution of the dust in the disk is defined by the following parameters: the inner and outer radius (R$_{\rm in}$ and R$_{\rm out}$) as well as the gap location R$_{\rm gap,in}$ and R$_{\rm gap,out}$; the total dust mass (M$_{\rm dust}$) and its distribution across the disk, the surface density profile (SDP); and finally an additional dust depletion factor for the inner disk f$_{\rm inner}$. The stellar spectrum is described by a Kurucz model with effective temperature T$_{\rm eff}$, luminosity L$_*$\footnotemark and mass, and is set at a distance d. These parameters are summarized in Table \ref{tab:model}. 

      Although the stellar parameters have recently been updated by \cite{2011arXiv1104.0905T}, our fit parameters provide an equally good fit to the stellar photosphere. Adopting the lower luminosity of 26 L$_{\odot}$ would have only a small impact on the model, as it requires a higher continuum opacity in the optical to achieve the same dust temperatures and fit the SED.

 \footnotetext{The value derived by \cite{1997A&A...324L..33V} is L$_*$=33 L$_{\rm \odot}$, but L$_*$=36 L$_{\rm \odot}$, which is within the 1$\sigma$ error, better fits the available photometry.}

      \subsection{Dust model}\label{sec:dust}
      We compute the optical properties of our grains using a distribution of hollow spheres \citep[DHS,][]{2005A&A...432..909M}, which has been shown to represent mid-infrared feature shapes very well \citep{2010ApJ...721..431J}. 

      We will restrict ourselves to a model with three dust components that dominate the opacity from the optical to the far-infrared: amorphous silicates, crystalline forsterite and carbon. Amorphous silicates produce the 10 and 20 \um emission features which are very prominent in HD 100546. We will use the composition and optical constants of interstellar silicates towards the galactic center as derived by \cite{2007A&A...462..667M}, which are summarized in table \ref{tab:min}. Crystalline forsterite dominates the mid infrared opacity only at wavelengths where it has strong features, most notably at 11.3, 23.5, 33.5 and 69 \um. We will use the optical constants from \cite{2006MNRAS.370.1599S}, as these have been measured at different temperatures. At optical and near-infrared wavelengths, the opacity is dominated by continuum sources without spectral features in the mid and far-infrared. For this component we use carbon, with optical constants from \cite{1993A&A...279..577P}. Note that probably not all of the continuum-opacity sources are amorphous carbon, and that also different materials with similar optical properties could contribute, for example metallic iron.

      \begin{table}
        \centering

        \title{Dust mineralogy}

        \begin{tabular}{llll}
          \hline \hline
          \multicolumn{2}{l}{Dust species} & reference \\
          \hline
                     & 13.8\% MgFeSiO$_{4}$ & [1] \\
          Amorphous  & 38.3\% Mg$_{2}$SiO$_{4}$ & [2] \\
          silicates  & 42.9\% MgSiO$_{3}$ & [1]\\
                     & 1.8\%  NaAlSi$_{2}$O$_{6}$ & [3]\\
          \hline
          Carbon     & C & [4] \\
          Forsterite & Mg$_{2}$SiO$_{4}$  & [5] \\
          \hline \hline
        \end{tabular}
        \caption{Dust composition used in this paper. Opacities are calculated from the optical constants using a size distribution from 0.1 to 1.5 \um proportional to a$^{-3.5}$. The shape of the particles is DHS, with f$_{\rm max}$= 0.7 except for crystalline forsterite, which has f$_{\rm max}$= 1.0.
          References to the optical constants:
          [1] \cite{1995A&A...300..503D};
          [2] \cite{1996A&A...311..291H};
          [3] \cite{1998A&A...333..188M};
          [4] \cite{1993A&A...279..577P};
          [5] \cite{2006MNRAS.370.1599S} \label{tab:min}
        }
      \end{table}

      Dust grains in protoplanetary disks are most likely in the form of mixed aggregates. Lacking a good effective medium theory for DHS, we calculate the opacities for every dust species from the optical constants as if they were separate particles, and then force them to be in thermal contact. Although this approach does not have as good a physical basis as an effective medium theory, it does allow an accurate representation of feature shapes. In addition, opacities do not have to be recalculated when the composition changes, making it easier to conduct large parameter studies and make models with spatially varying dust compositions, including temperature dependent opacities. The implications of the dust grains being in thermal contact will be discussed in section \ref{sec:tcontact}.
      
      Other dust properties - grain size range and composition - have to be constrained from the observations and are discussed in section \ref{sec:min}.

      \section{Disk structure}\label{sec:struct}
      Before we can model the spatial distribution of forsterite, we need to construct a structural model for the disk that fits the available observations. The disk of \object{HD 100546} has been well-studied, and the disk geometry is already well-constrained. In the next section we will describe these observational constraints. In section \ref{sec:SED} we fit the SED to fill in most of the remaining parameters of our disk model. The only parameter important for this study that can not be well constrained from the SED is the surface density profile (SDP). We will assume the same SDP as previous modeling attempts, which is consistent with VISIR imaging (see appendix \ref{sec:SDP}). Fig \ref{fig:image} shows a Q band image of the final disk model, illustrating its geometry with an outer disk (Section \ref{sec:outer}), gap (Section \ref{sec:gap}) and inner disk (Section \ref{sec:inner}).

      \begin{table}
        \centering
        \title{Model parameters}\\
        \begin{tabular}{lll}
          \hline \hline
          Parameter  & Value & reference \\
          \hline  
          T$_{\rm eff}$ [K]      & 10500  & [1] \\
          L$_*$ [L$_{\rm \odot}$]  & 36\fit  & [1] \\
          M$_*$ [M$_{\rm \odot}$]  & 2.4   & [1] \\
          d [pc] & 103 & [1] \\
          \hline  
          R$_{\rm in}$ [AU]  & 0.25  & [2] \\
          R$_{\rm exp}$ [AU] & 350 & [3] \\
          R$_{\rm out}$ [AU] & 1000 & [4] \\
          R$_{\rm gap,in}$ [AU] & 4 & [2] \\
          R$_{\rm gap,out}$ [AU] & 13 & [5] \\
          \hline 
          M$_{\rm dust}$ [M$_{\rm \odot}$] & 1\e{-4} \fit &  \\
          $\Sigma(r)$ & r$^{-1}$ \fit &  \\
          f$_{\rm inner}$ &  200\fit &  \\
          \hline 
          a$_{\rm min}$[\um] & 0.1 & [6] \\
          a$_{\rm max}$[\um] & 1.5 & [6] \\
          carbon fraction [\%] & 5 \fit &  \\
          Shape  & irregular (DHS) & [6] \\
          \hline 
          i [\degr] & 42 & [7] \\
          PA & 145 & [7] \\
          \hline \hline
        \end{tabular}
        \caption{Model parameters. All parameters with a dagger (\fit) are (re)fitted.
          References:
          [1] \cite{1997A&A...324L..33V};
          [2] \cite{2010A&A...511A..75B};
          [3] \cite{2010A&A...519A.110P};
          [4] \cite{2007ApJ...665..512A};
          [5] \cite{2005ApJ...620..470G};
          [6] \cite{2010ApJ...721..431J};
          [7] \cite{2000A&A...361L...9P}
          \label{tab:model}}
      \end{table}

      \subsection{Observational constraints}
      \begin{figure}
        \includegraphics[width=\linewidth]{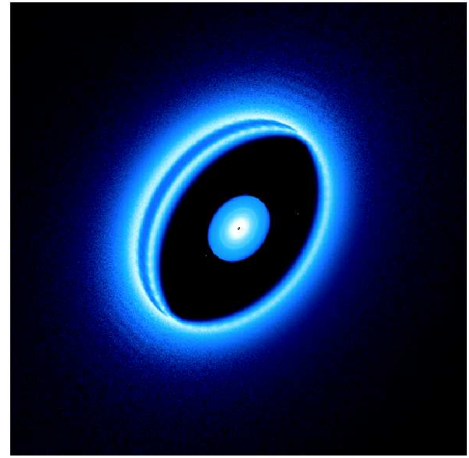}
        \caption[]{18.7 \um model image using our final model of the inner regions of HD 100546. Note that the inner disk casts a clear shadow on the outer disk wall. The field of view of the image is 0.5\arcsec{} by 0.5\arcsec, or 51.5 AU at a distance of 103 pc. North is up and East is left.
          \label{fig:image}}
      \end{figure}

      \subsubsection{Outer disk}\label{sec:outer}
      The (outer) disk of HD 100546 was first resolved in scattered light by \cite{2001A&A...365...78A}, and has since then been imaged at different wavelengths (e.g. \citealt{2003ApJ...598L.111L}, \citealt{2007ApJ...665..512A}). Its position angle and inclination have been measured using different techniques, indicating an inclination between 40 and 50 degrees from face-on, and a position angle east of north between 130 and 145 degrees \citep[e.g.][]{2000A&A...361L...9P, 2007ApJ...665..512A}. We adopt the inclination of 42 degrees and a position angle of 145 degrees.

      The outer radius of the dust disk is difficult to determine. Scattered light is observed up to at least ten arcseconds (1000 AU at a distance of 103 pc, \citealt{2007ApJ...665..512A}), yet it is not clear whether this belongs to the disk or to a remnant envelope, and the dust disk could be significantly smaller \citep[e.g.][]{2007ApJ...665..512A}. Modelling of rotational CO lines are consistent with a gas-rich disk with an outer radius of 400 AU \citep{2010A&A...519A.110P}. We therefore adopt an outer radius of 1000 AU, but with an exponential cutoff that sets in at R$_{\rm exp}$= 350 AU \citep{2008ApJ...678.1119H}.

      The total dust mass of the outer disk is 7.2\e{-4} M$_{\odot}$, as has been determined by \cite{1998A&A...336..565H} from 1.3 mm measurements, which probe mostly big grains in the disk midplane. The mass distribution in the outer disk is not well constrained. Various authors have adopted an SDP of $\Sigma(r)=r^{-p}$ with p=1 (e.g. \citealt{2010A&A...519A.110P}, \citealt{2010A&A...511A..75B}), a value commonly used for protoplanetary disks. We will also adopt this value, which is consistent with our VISIR image in appendix \ref{sec:SDP}

      \subsubsection{Gap size}\label{sec:gap}

      \begin{figure}
        \includegraphics[width=\linewidth]{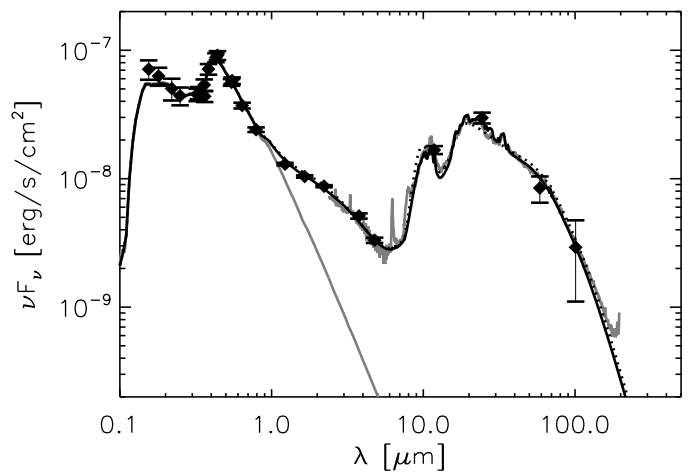}

        \caption[]{SED of HD 100546 (diamonds). The grey line is the ISO spectrum. Overplotted is the best-fit disk model with forsterite (R20, solid line) and without forsterite (dotted line), both without PAHs. \label{fig:sed}}
      \end{figure}

      The SED of HD 100546 (Fig. \ref{fig:sed}) is dominated by a relatively small near-infrared excess between 2 and 8 \um, followed by a steep rise towards the mid-infrared at 10 \um. This SED has been attributed to a disk gap around 10 AU - possibly cleared out by a proto-planet - which creates a frontally illuminated wall at the far edge of the gap \citep{2003A&A...401..577B}. This wall has a temperature of 200 K, and is responsible for the huge mid-infrared excess compared to other Herbig Ae/Be stars. The gap manifests itself in mid-infrared imaging, where the source is much larger than other Herbig Ae/Be stars, and its size is independent of wavelength \citep{2003ApJ...598L.111L}.

      The gap is also seen in the distribution of the circumstellar gas (e.g. \citealt{2006A&A...449..267A}, \citealt{2009ApJ...702...85B},\citealt{2009A&A...500.1137V}), and tight constraints have been put on the size of the disk gap using spatially resolved spectroscopy by \cite{2005ApJ...620..470G}. We adopt their value for the gap outer radius of R$_{\rm gap,out}$=13 AU, which is the location of the disk wall. 

      \subsubsection{Inner disk}\label{sec:inner}
      The inner disk of HD 100546 causes a near-infrared excess that is relatively small compared to other Herbig Ae/Be stars \citep{2003A&A...401..577B}, indicating that the dust is depleted \citep{2003A&A...398..607D} or settled. \cite{2010A&A...511A..75B} have resolved the disk using near-infrared interferometry, constraining the inner radius to R$_{\rm in}$=0.25 AU. The authors also find that the inner disk is very tenuous, and has a low dust mass of only 3\e{-10} M$_\odot$. The AMBER observations cannot resolve the outer radius, but they put it at R$_{\rm gap,in}$=4 AU. We adopt these values for the inner and outer radius.

      \subsection{Mineralogy}\label{sec:min}
      Before determining the remaining model parameters using SED fitting we need to specify our dust composition. Although the peculiar forsterite mineralogy of HD 100546 is the main focus of this paper, forsterite itself plays only a minor role in determining the disk structure and fitting the SED. The main reason for this is that it has a very low opacity in the optical compared to other dust species such as amorphous silicates and carbon. The latter species therefore determine the heating/cooling balance of the dust, and set the disk's temperature and vertical density structure. Leaving out the forsterite does not affect the shape of the modeled Q band radial profile, and changes to the overall SED are small (Figure \ref{fig:sed}). The shape of the 10 micron feature is better fitted with forsterite included, but this does not affect the overall SED shape (see Figure \ref{fig:sed}). We will therefore constrain the disk geometry in this section using only the other two components, amorphous silicates and carbon. By doing so, the derived geometry is independent of the forsterite mineralogy, and we can study the abundance and spatial distribution of forsterite in section \ref{sec:fors} without refitting the SED.

      The dust composition is determined by the mass fraction of carbon, and follows from our SED fit in section \ref{sec:SED}. Both \cite{2003A&A...401..577B} and \cite{2010ApJ...721..431J} show that the small grains responsible for the mid-infrared dust emission features in HD 100546 cannot be larger than a few \um, though larger grains could be present at higher optical depths not probed by Spitzer. The small grains are most important for setting the disk density and temperature structure because of their high opacities \citep{2008A&A...492..451M}, and we adopt a grain size distribution between 0.1 and 1.5 \um for all grains. We do not fit the SED in the (sub)mm because it probes a separate population of large, millimeter-sized grains \citep{2010A&A...511A..75B}, which do not influence the strength of mid-infrared features. A summary of the mineralogy can be found in table \ref{tab:min}.

      \subsection{SED fitting}\label{sec:SED}
      With the disks radii (R$_{\rm in}$, R$_{\rm gap,in}$, R$_{\rm gap,out}$, R$_{\rm out}$) and opacities constrained, the next step is to constrain the dust mass (M$_{\rm disk}$), inner disk depletion (f$_{\rm inner}$) and carbon fraction by SED fitting. The SED is relatively insensitive to the SDP, but some limits can be placed on it using spatially resolved VISIR imaging (See appendix \ref{sec:visir}). For now, we will adopt $p=1$. The total dust mass for HD 100546 has been measured at millimeter wavelengths \citep{1998A&A...336..565H}. However this mass probes mostly the big, millimeter sized grains in the midplane, while we are interested in the dust mass in small micron-sized grains responsible for mid-infrared  and far-infrared features. We therefore have to refit the dust mass to the optically thin part of the SED at \ong60-200 \um. We find a dust mass of \ten{-4} M$_{\odot}$, consistent with the dust mass in small grains from \cite{2003A&A...398..607D} and \cite{2010A&A...511A..75B}.

      We proceed by fitting the inner disk (0.25-4 AU), which has been done before with a similar 2D radiative transfer code by \cite{2010A&A...511A..75B}. They use a parametrized vertical structure with a fixed scale height and flaring exponent. In our model, the vertical structure is computed under the assumption of vertical hydrostatic equilibrium. Therefore we can not freely adjust these parameters, so instead we fit the inner disk SED by varying its dust mass. We do this by assuming the mass distribution in the inner disk follows the same power law as the outer disk ($p=1$) - as would be the case for a disk without a gap - and reduce the surface density\footnote{Reducing the surface density has a similar effect on the near-infrared flux as reducing the scale height or increasing the flaring exponent.} to lower the optical depth and fit the near-infrared flux in the same way as \cite{2010A&A...511A..75B}. We fit a dust mass of 6\e{-9} M$_\odot$, consistent with the value found by \cite{2010A&A...511A..75B}. This corresponds to a surface density normalization of $\Sigma_{1\rm AU}=$ 2.3\e{-3} g/cm$^2$ . This is a reduction of a factor f$_{\rm inner}$= 200 with respect to an extrapolation inward of the SDP of the outer disk, which has $\Sigma_{1\rm AU}=$ 0.45 g/cm$^2$ .

      The modelled inner disk is not completely optically thin in the radial direction at optical wavelengths, as can be inferred from the shadow seen on the disk wall in figure \ref{fig:image}. Even when we take into account that the star is not a point source \citep{2010ApJ...717..441E}, the shadow reaches up to a height above the midplane of around $\sim$ 1 AU at the disk wall. The shadow only disappears when the inner disk has an increased scale height such that it effectively becomes a halo \citep{2010A&A...512A..11M, 2011arXiv1101.5719V}, as has been suggested for HD 100546 by \cite{2006ApJ...636..348V}.

      Even though the wall at 13 AU is partly shadowed, the illuminated part still intercepts a significant fraction of the stellar light. It is frontally illuminated by the star, and has a temperature of 200 K. Such a structure fits the mid-infrared part of the SED without any increase in scale height beyond hydrostatic equilibrium - which is often needed to account for extreme near-infrared excesses in Herbig Ae/Be stars \citep{2009A&A...502L..17A}. The steep rise in the SED between 8 and 10 \um is caused by this wall, but it also contains information on the dust composition. At 8 \um the opacity is dominated by carbon, at 10 \um by amorphous silicates. From the 10/8 \um flux ratio, we estimate a carbon/silicate mass ratio of 5\%. Increasing the carbon fraction 'fills' the gap in the SED at 8 \um: with 10\% carbon the flux at this wavelength is already significantly overpredicted. Because carbon combusts (see, e.g., \citealt{2010ApJ...710L..21L}) at the same temperatures needed to anneal silicates \citep{2001A&A...378..192G}, the relatively low carbon abundance could be linked to the high crystallinity.

      \subsection{Final model}\label{sec:final}
      Model parameters of the final disk model used for the analysis of the forsterite features can be found in table \ref{tab:model}. Note that this model is in hydrostatic equilibrium, and therefore does not have a predefined scale height. The calculated scale height increases with height above the midplane due to the temperature gradient, but can be fitted with a power law in radius (H$= \rm H_{100AU} * (r/100 AU)^{\beta}$) at both the surface and midplane for comparison. The outer disk is well described by a flaring exponent of $\beta$ = 1.4 and a scale height that increases from H$_{\rm 100AU}$=13 AU in the midplane to H$_{\rm 100AU}$=21 AU at the disk surface and in the disk wall. The inner disk is fitted with $\beta$ = 1.3 and H$_{\rm 100AU}$=13 AU in the midplane to H$_{\rm 100AU}$=25 AU in the disk rim. The midplane temperature follows T = $\sim$150 * (r/AU)$^{-0.25}$ for the outer disk, and is a factor of 2 higher in the inner disk.

      \section{Disk mineralogical model:  Forsterite}\label{sec:fors}
      In this section we will study the abundance and spatial distribution of forsterite in the disk of \object{HD 100546}, using the disk structural model described in the previous section. The aim is to distinguish between the two scenarios for the forsterite composition described by \cite{2010A&A...518L.129S} (warm iron-free forsterite or cold iron-containing forsterite), which have also been discussed in the introduction.

      \subsection{Optical depth effects}\label{sec:opticaldepth}

      \begin{figure}
        \includegraphics[width=\linewidth]{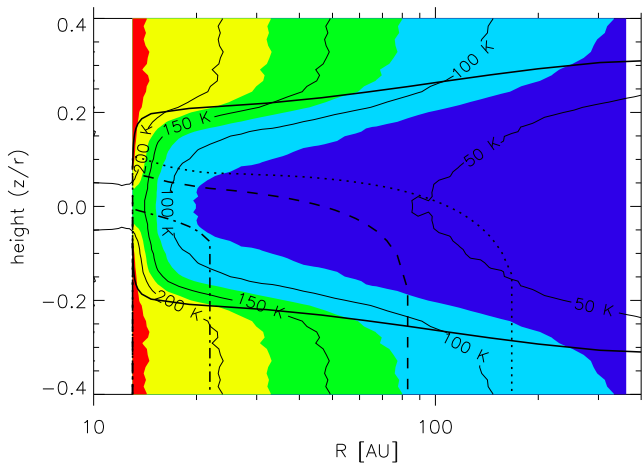}
        \caption[]{The temperature structure of the outer disk of HD 100546. Overplotted are the radial $\tau$=1 surface in the optical (solid thick line) as well the vertical $\tau$=1 surfaces of the continuum at 24, 33 and 69 \um (dotted, dashed, dot-dashed). Note that the vertical $\tau$=1 surface is calculated from above, meaning that regions with higher optical depth lie below it. The 69 \um surface lies below the disk midplane, meaning that the disk is close to getting optically thin at these wavelengths.
          \label{fig:temp}}
      \end{figure}

      \begin{figure}
        \includegraphics[width=\linewidth]{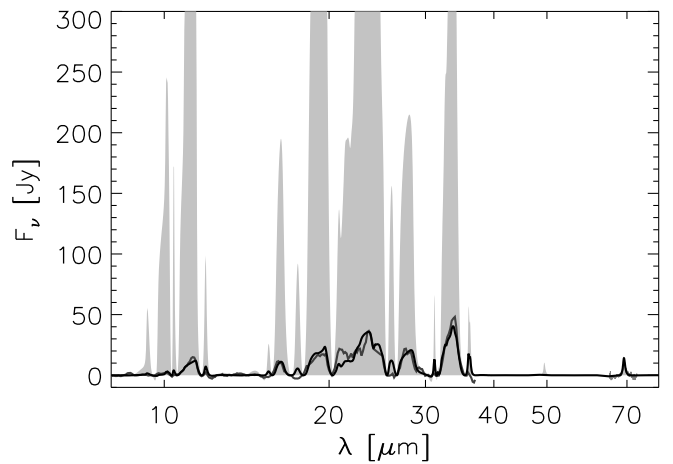}
        \caption[]{Continuum subtracted spectra comparing the optically thick (solid black) and optically thin (light grey area) approach using only warm forsterite (model RW30). Models are fitted to the 69 \um feature shape. Overplotted are the Spitzer spectrum between 7 and 35 \um and the PACS spectrum between 65 and 73 \um (dark grey lines). A more detailed version comparing the optically thick model and observations can also be found in figure \ref{fig:para}, panel RW30.
          \label{fig:cont_sub}}
      \end{figure}

      We explore in detail the effects of optical depth and the gapped disk geometry of HD 100546 on the observed feature strengths of forsterite. Throughout the paper, we will assume that the forsterite is in thermal contact with the other dust species (see section \ref{sec:min}). Fig \ref{fig:temp} shows the temperature structure in the outer disk from the model derived in section \ref{sec:final}. Fig \ref{fig:temp} also shows the vertical $\tau$=1 surfaces - as seen from the observer to the emission point - of the continuum at 24, 33 and 69 \um. These surfaces mark the boundaries below which the dust can no longer contribute substantially to the feature at that particular wavelength.
      
      At 69 \um, the vertical $\tau$=1 surface lies below the disk midplane (i.e., the disk becomes optically thin), allowing most of the disk to contribute to the 69 \um feature. At shorter wavelengths, these surfaces lie above the disk midplane out to radii of 50-100 AU. For the 150 and 200 K components, a significant part lies below the $\tau$=1 surface, especially when taking into account that the density increases towards the midplane. Most of the warm forsterite that can contribute to the 69 \um feature is therefore hidden from view at shorter wavelengths.

      To illustrate the influence of these optical depth effects on the measured feature strengths of forsterite, we compare the continuum subtracted spectra (see appendix \ref{sec:cont_sub}) of our model (described in section \ref{sec:only_warm}) with the optically thin model from \cite{2010A&A...518L.129S} in Figure \ref{fig:cont_sub}. Both models are fitted to the feature shape and strength at 69 \um, and contain only 150 K and 200 K forsterite. The optically thin approach clearly overpredicts the strength of the short wavelength features in the 10 to 40 \um range, while the 2D radiative transfer does not overpredict these because the bulk of the warm forsterite is hidden at such short wavelengths.

      The gapped geometry of HD 100546 plays a key role here: because part of the wall at 13 AU is directly illuminated by the central star, it heats regions close to the midplane to temperatures of 150 to 200 K. In a disk without a gap, these regions would be much colder and these high temperature regions would only be found in the midplane at much smaller radii ($\sim$4 AU). There, the surface density would be high enough also for the $\tau$=1 surface at 69 \um to be above the midplane. Without a gap, the optical depth effects would not be so pronounced as the hot forsterite would be hidden from view at 69 \um as well.

      This first analysis shows that while optically thin models with iron-free forsterite overpredict feature strengths at shorter wavelengths, a 2D radiative transfer model that takes optical depth effects into account does not. Additionally, none of the longer wavelength features (33, 69 \um) require an additional cold ($<$150 K) component to fit the feature strengths. This indicates that the shape of the 69 \um feature can be fitted with only warm forsterite, suggesting an iron-free composition (the first scenario from \citealt{2010A&A...518L.129S}). We will therefore first pursue a disk model that explains all feature strengths and shapes with iron-free forsterite. Afterwards, we will discuss models that include a fraction of iron. We describe how we implement the temperature-dependent opacities of iron-free forsterite into our radiative transfer code in Appendix \ref{sec:topac}.

      \subsection{Forsterite spatial distribution}
      To determine the spatial distribution of the forsterite in the outer disk, we use the disk structural model from section \ref{sec:final}. We replace a small fraction of the amorphous silicates by forsterite, and compare continuum subtracted model spectra with the data. All discussed models can be found in Fig. \ref{fig:para}. These models are fitted to the integrated flux at 69 \um, which is $\sim$9 Jy \um.

      We will show that models with constant abundance throughout the outer disk severely underpredict short wavelength features due to optical depth effects (model G0). We therefore run an extensive parameter study where the radial distribution of forsterite peaks towards the disk wall. We consider two parametrizations: an abundance gradient (model series G), predicted by radial mixing models (\citealt{2001A&A...378..192G}), and a (narrow) ring of constant abundance (model series R), which makes it easier to control the minimum temperature of the forsterite. In addition, we also ran a series of models where the forsterite is only located in the disk wall and at the disk surface (model RW), and not in the midplane. Below we discuss the different models.

      \subsubsection{Constant abundance throughout the disk}
\begin{figure*}
        \includegraphics[width=0.79\textwidth]{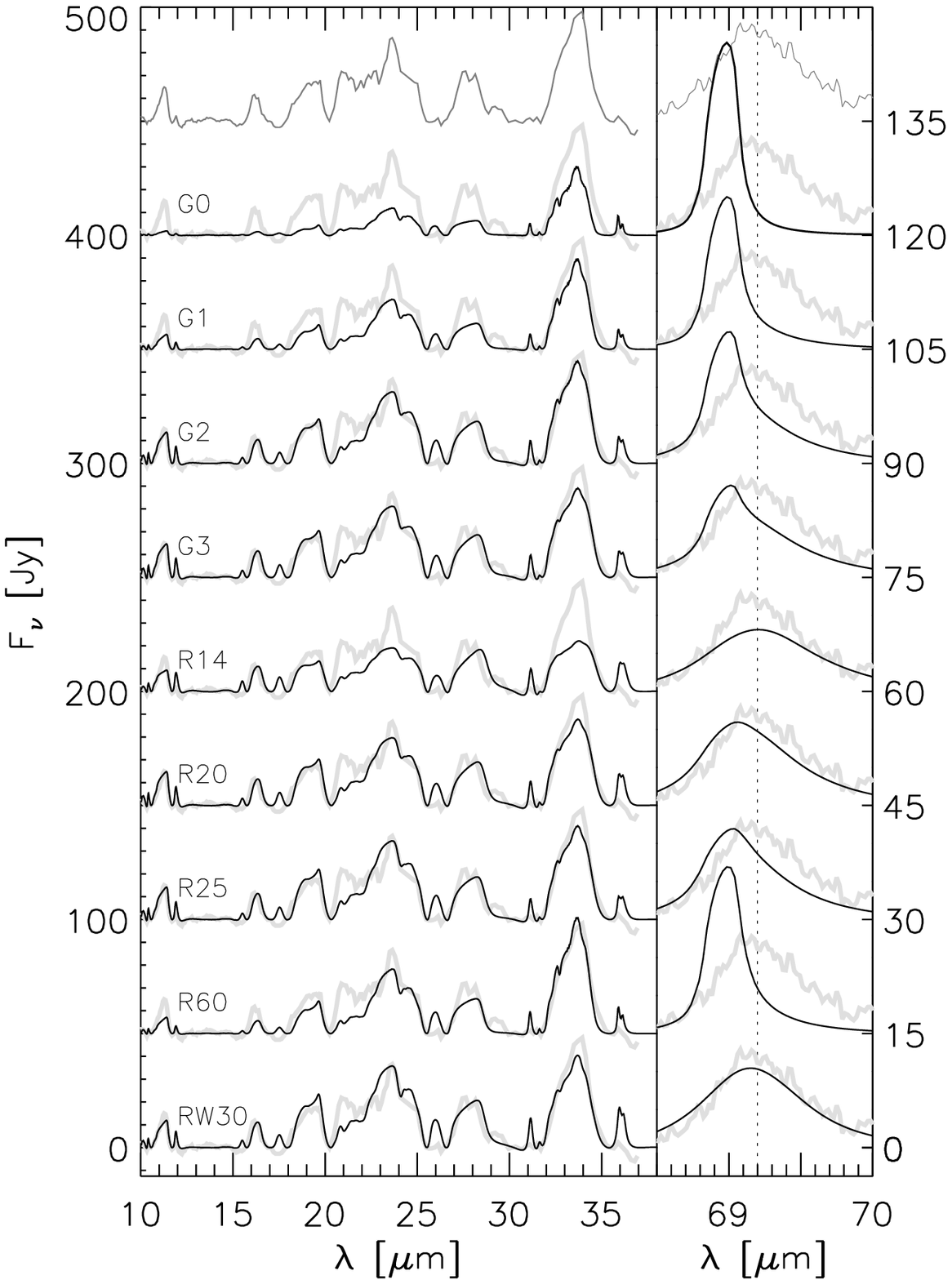}
        \hspace{0.005\textwidth}
        \includegraphics[width=0.195\textwidth]{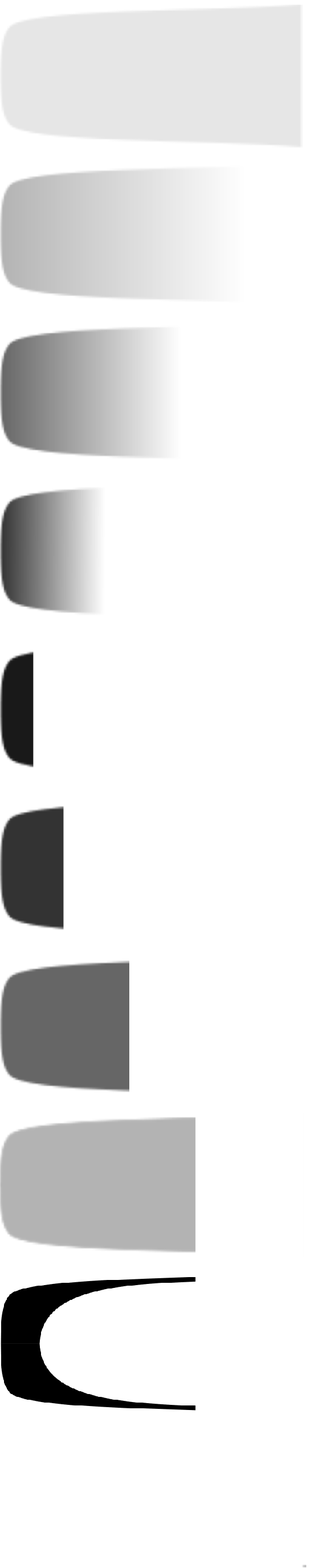}
        \caption[]{Continuum subtracted spectra comparing the models (solid black) and Spitzer and PACS observations (grey). Presented models have integrated feature strength between 67.5 and 71.0 \um of 8.5 $\pm$ 1.5 Jy. Offsets between models are 50 and 15 Jy for the left and right panel respectively. The cartoons on the right show the location of the forsterite in the outer disk, not to scale. A higher intensity represents a higher abundance.
Roman symbols present models with:
\\G0   - 2\% forsterite throughout the entire disk (gradient with r$^{-0}$);
\\G1   - gradient with 10\% forsterite at 13 AU and r$^{-1}$;
\\G2   - gradient with 30\% forsterite at 13 AU and r$^{-2}$;
\\G3   - gradient with 40\% forsterite at 13 AU and r$^{-3}$;
\\R14  - ring with 60\% forsterite between 13 and 14 AU;
\\R20  - ring with 40\% forsterite between 13 and 20 AU;
\\R25  - ring with 30\% forsterite between 13 and 25 AU;
\\R60  - ring with 10\% forsterite between 13 and 60 AU;
\\RW30 - only 150 and 200 K forsterite, 60\% between 13 and 30 AU;

          \label{fig:para}}
      \end{figure*}

      In contrast with the result from \cite{2003A&A...401..577B}, models with a constant forsterite abundance throughout the outer disk consistently fail to reproduce all observed feature strengths. This failure is due to optical depth effects, which suppress short wavelength features (Section \ref{sec:opticaldepth}). A very low abundance of 2 \% fits the integrated 69 \um flux (Fig. \ref{fig:para}, model G0), but underpredicts shorter wavelength features, most notably shortwards of 33 \um. An abundance increase of at least a factor of 10 is necessary in the disk wall to fit the flux at 11 \um. To achieve this without overpredicting the 69 \um feature by the same factor, we need to concentrate the forsterite towards the disk wall.

      \subsubsection{Abundance gradient}
      First, we concentrate the forsterite towards the disk wall by introducing an abundance gradient that gradually decreases outward. Radial mixing models predict that if radial mixing transports crystalline material from the inner to the outer disk, it follows a smooth distribution  \citep{2001A&A...378..192G}, which is well described by a power law \citep{1990ApJ...348..730S}. We parametrize the forsterite abundance as $f(r) = f_{\rm 13 AU}~(r[AU]/13 AU)^{-\xi}$, where free fitting parameters are the abundance in the disk wall (f$_{\rm 13 AU}$) and the steepness of the gradient ($\xi$). The grid scanned for these parameters is summarized in table \ref{tab:forsterite}. We start by assuming the forsterite is well-mixed in the vertical direction, i.e. there is no vertical abundance gradient.

      With a gradient of $\xi$=1, we can increase the forsterite abundance in the wall to 10\% without overpredicting the 69 \um flux. This is still not high enough to explain the short wavelength features, which originate mainly in the disk wall (G1). Steeper gradients of $\xi$=2 or 3 concentrate the forsterite closer to the disk wall, and can reach higher abundances for the same 69 \um flux. They reach abundances of respectively 30\% (G2) and 40\% (G3), which is sufficient to fit the feature strengths at shorter wavelengths. With even steeper gradients, the abundance drops off so rapidly even in the wall that short wavelength features are again underpredicted (not plotted).

      Although models G2 and G3 can explain all feature strengths, the shape of the 69 \um feature does not fit the observations well. The peak is located at 69.0 \um, indicating that cold, 100 K forsterite dominates the peak location. Although an increasing contribution at 69.2 \um is seen with steeper gradients, it never dominates the feature shape. These solutions therefore require some iron to shift the feature towards longer wavelengths. However, even then the feature shapes do not match very well (Figure \ref{fig:para}).

      The reason why the cold forsterite is so prominent at 69 \um is that even with a steep gradient a few percent forsterite remains far out in the disk. Such a low abundance provides a significant contribution at 69 \um, while it does not add a lot of flux to the short wavelength features (see also model G0). We will therefore focus on a series of models with a forsterite abundance that is non-zero only near the disk wall.

      \begin{table}
        \centering
        \title{Forsterite spatial distribution}
        \begin{tabular}{lll}
          \hline \hline
          Parameter  & Best-fit & Range explored\\
          \hline
          abundance [\%]      & 40 & \{2, 5, 10, 20, 30, 40, 50, 60, 70, 80, 90\} \\
          r$_{\rm ring}$ [AU]  & 20 & \{14, 15, 16, 17, 18, 19, 20, 25, \\
          &    &  \hspace{0.5cm} 30, 35, 40, 60, 80, 100, 400, 1000\} \\
          $\xi$               & 2   & \{0, 0.5, 1.0, 2.0, 3.0, 4.0\} \\
          
          \hline \hline
        \end{tabular}
        \caption{Forsterite abundances in the disk. Best-fit and ranges explored. Abundance refers to abundance at 13 AU; outwards the abundance is either constant up to r$_{\rm ring}$, beyond which it is zero, or it drops off gradually as $\big(\frac{r[AU]}{13 AU}\big)^{-\xi}$  \label{tab:forsterite}}
      \end{table}

      \subsubsection{Abundance ring}
      To prevent a low percentage of cold forsterite in the outer disk from dominating the shape at 69 \um, we choose a second parametrization where the forsterite is located in a narrow ring starting at the disk wall. The abundance of forsterite is constant within the ring, between 13 and r$_{\rm ring}$ AU, and is zero outside this ring. Both the abundance in the ring and its outer radius r$_{\rm ring}$ are free fitting parameters. The explored parameter space is summarized in Table \ref{tab:forsterite}, a selection of models is displayed in Figure \ref{fig:para}.

      Because the midplane is close to optically thin at 69 \um, avoiding a strong contribution from cold (50-100 K) forsterite requires a narrow ring of only a few AU. A ring with outer radius of 14 AU (R14) requires a high abundance of 60\% to fit the integrated feature strength at 69 \um. At such high abundances, features become optically thick and flatten off, producing a 69 \um feature that is much too broad. Additionally, the short wavelength features flatten off, and become too weak compared to the observations. 

      Lower abundances require broader rings to produce enough emission at all wavelengths. Rings with abundances of 40 and 30\% require outer radii of 20 (R20) and 25 AU (R25) respectively. At these radii, midplane temperatures are down to 100 K (Fig. \ref{fig:para}), and the 69 \um feature is dominated by this component with a peak at 69 \um. In contrast to the gradient models G1 and G2, these rings do not contain a 50 K component, since these temperatures occur farther out in the disk. Therefore the feature shape matches much better than models with a gradient, although still some iron is required to shift the feature towards longer wavelength. At even lower abundances (R60, 10 \% and 60 AU), the feature strengths at shorter wavelengths cannot be reproduced.

      \subsubsection{Only warm forsterite}\label{sec:only_warm}
      In order to reproduce the 69 \um feature with iron-free forsterite, we need to exclude contributions from 100 K forsterite. However, ring models that do so (R14) do not produce enough flux at all wavelengths because the emitting region is too small. Larger rings include warm forsterite from the disk surface, but also colder forsterite from the midplane (R20 and R25). We therefore run the same series of ring models, but now with forsterite only in regions that are 150 and 200 K, disabling the contribution from colder forsterite. This model is easily compared to the optically thin fit of \cite{2010A&A...518L.129S} (Fig. \ref{fig:cont_sub}), which also has only these components. We implement this by replacing the opacities for 50, 100 and 295 K forsterite with that of amorphous silicates. A ring with an outer radius of 30 AU and abundance of 60 \% provides a good fit to the short wavelength features (RW30). The shape of the 69 \um feature matches the observed one well.

      Looking at figure \ref{fig:temp}, it becomes clear that such a model with only warm forsterite has an abundance that varies with height above the midplane. Outside of the disk wall ($>15$ AU), a vertical temperature gradient is present from a warm surface to a cold midplane. The forsterite in this region must therefore be located only at the disk surface, with an abundance that decreases to zero near the disks midplane. The forsterite is present in the disk wall and near the disk surface farther out, but \textit{not} in the midplane. This scenario agrees well with the geometry proposed by \cite{2003A&A...401..577B}, where forsterite is produced in the disk gap and transported into the disk wall and outer disk surface.

      \subsubsection{Best fit to 69 \um feature}

      \begin{figure*}
        \includegraphics[width=0.49\linewidth]{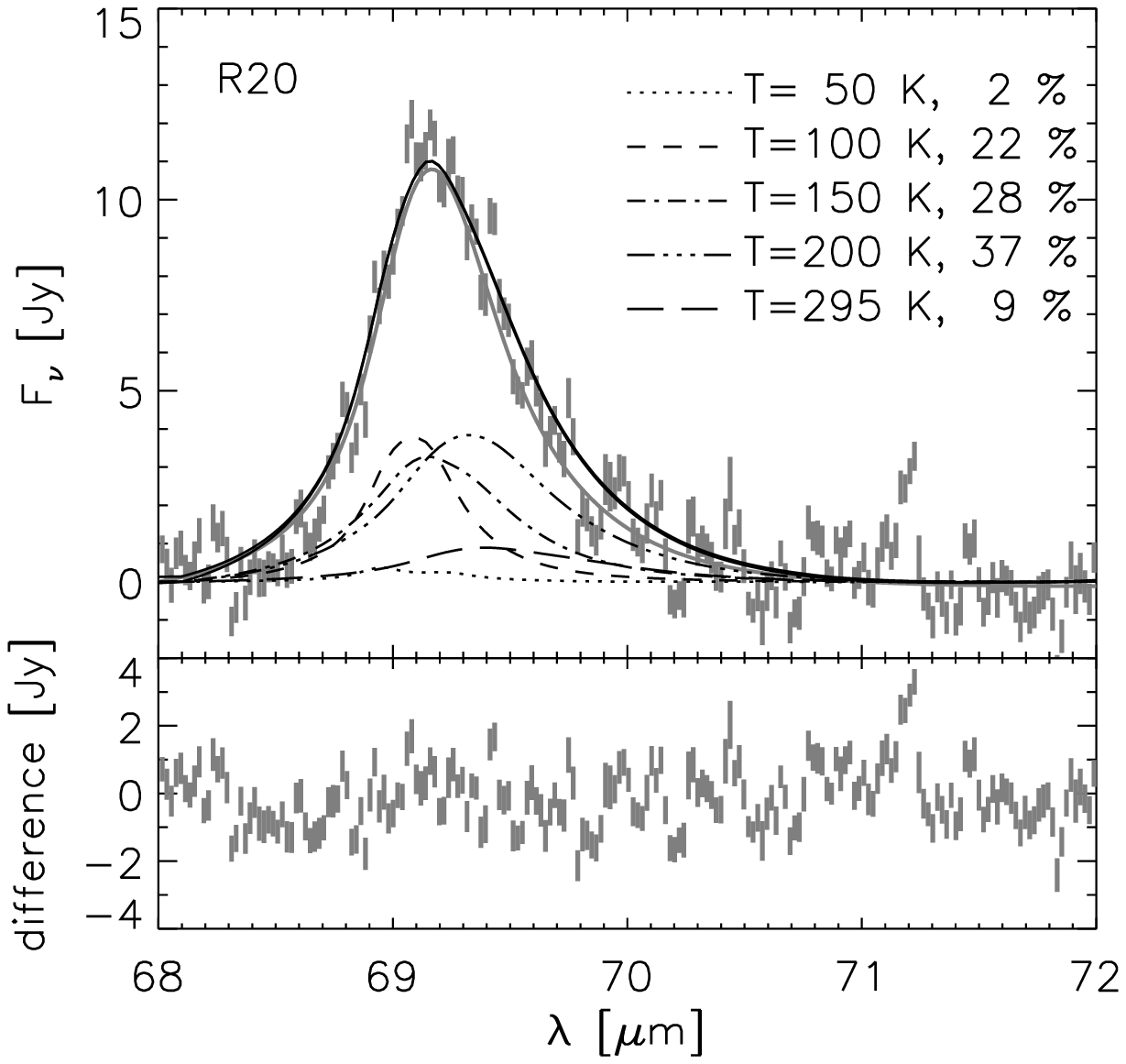}
        \includegraphics[width=0.49\linewidth]{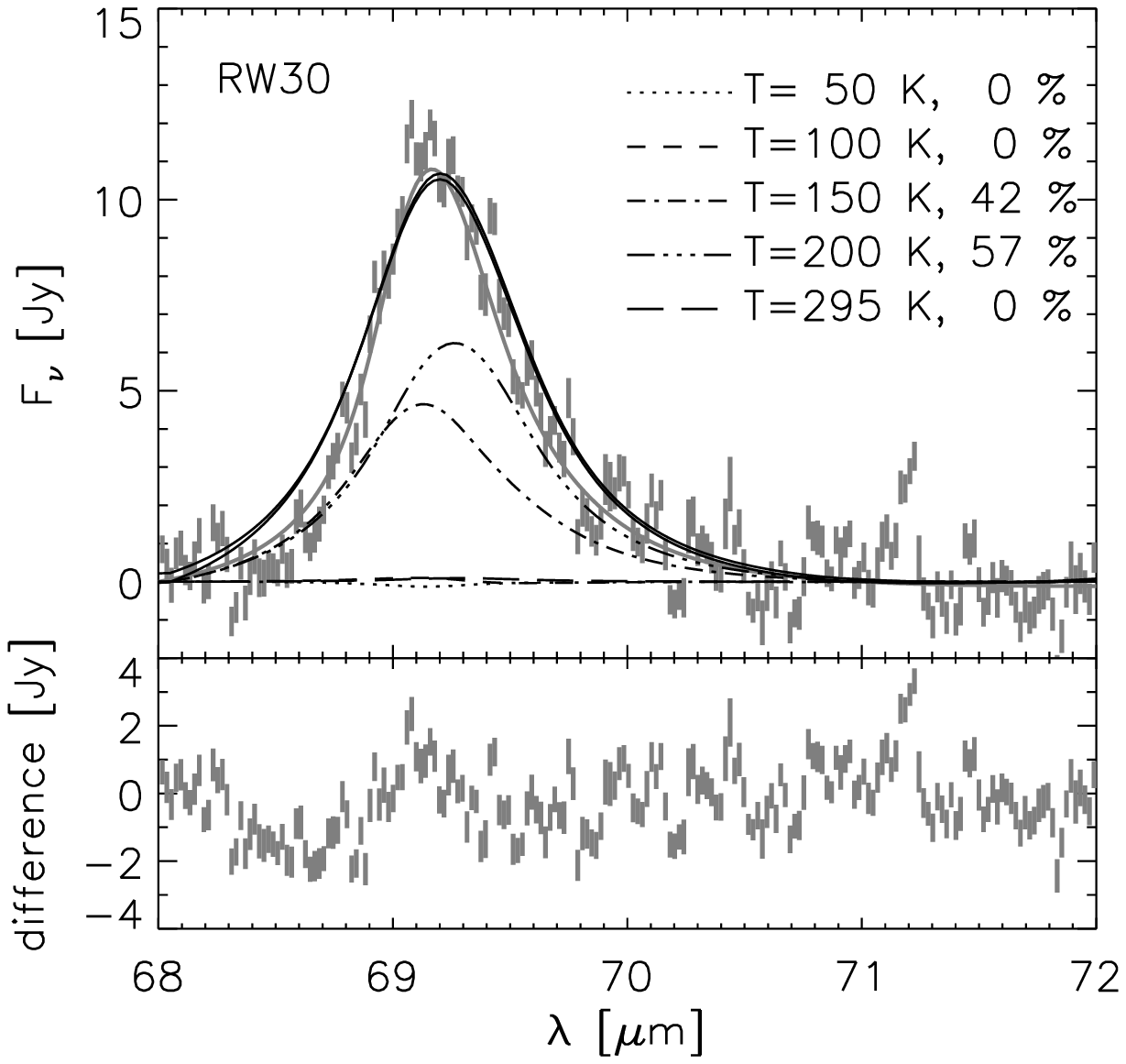}
        \caption[]{Modelled 69 \um feature shape (solid lines) compare to the PACS observations (grey error bars) for models R20 with a 0.1 \um shift towards longer wavelengths (left) and model RW30 with a 0.05 \um shift (right). Overplotted are the optically thin fit from \cite{2010A&A...518L.129S}(grey solid line), and contributions from the individual temperature components. The bottom panels show the difference between the PACS observations and the modelled spectra.
          \label{fig:fs69}}
      \end{figure*}

      The parameter study from the previous section yields a number of good fits to the integrated feature strengths, but only 2 models fit the shape of the 69 \um feature well: model R20, with 40\% forsterite between 13 and 20 AU and model RW30, with 60\% forsterite between 13 and 30 AU, but only in the 'warm' regions with a temperature of 150 and 200 K. Fig. \ref{fig:fs69} shows the modeled feature shape, as well as the contribution of the individual temperature components.
      
      Model R20 peaks at 69.10 \um, and the model has been shifted by 0.1 \um to obtain a good fit. We interpret this shift as an increase in the iron fraction from 0 to 0.3 \%, following the interpolation for low iron content by \cite{2010A&A...518L.129S}, which gives a peak shift of 3.13 \%/\um. The peak shows dominant contributions from both 200, 150 and 100 K, and is therefore slightly broader than the optically thin fit from \cite{2010A&A...518L.129S}, but well within the error bars. The higher temperature components dominate the feature by integrated emission, but the 100 K component is narrower and is equally important for the peak position.

      Model RW30 peaks at 69.15 \um, and provides a good by-eye fit without a wavelength shift. Fit quality increases with an additional shift of 0.05 \um, corresponding to an iron fraction of less than 0.2 \%. The relative contribution of the 200 and 150 K components are 57 and 42\% in contrast with 62 and 27 \% for the optically thin model (B. Sturm, private communication). This also explains the small wavelength shift required, as the 150 K component has its peak at 69.10 \um.

      The feature of model RW30 is also a little bit too broad, especially at the blue side of the feature. The reason for this broad feature is the high forsterite abundance of 60\%, at which the feature itself starts to become optically thick. A lower abundance would require a larger region with a temperature of 200 K, which is difficult to achieve in the current model.

      \subsection{Location of forsterite emission}\label{sec:location}
      To visualize where the different forsterite features originate in the disk, we calculate the contribution to the features at every location in the disk. Although the temperature and optical depth also give some information about where the features originate (Fig. \ref{fig:temp}), it does not take into account that the density decreases with disk height and radius, as well as the increase in opacity in the wavelength of the feature and inclination effects.

      To calculate these contributions, the ray tracer stores the individual contributions to the total spectrum in every grid cell. We then fit a 3rd order polynomial to the continuum, subtract it, and calculate the integrated feature strength. The contributions of the individual temperature components to the 69 \um feature are calculated in a similar fashion.

      Because we view the disk at a 42\degr{} angle, the upper half of the disk (z/r$>$0) is in our direct line of sight, while the bottom half (z/r$<$0) is mostly obscured from view. This results in an asymmetric profile around the midplane (z/r=0) in Figures \ref{fig:vs} and \ref{fig:vs69}, while the forsterite \textit{abundance} itself is symmetric around the midplane in all our models.

      \subsubsection{Short wavelength features}

      \begin{figure}
        \includegraphics[width=\linewidth]{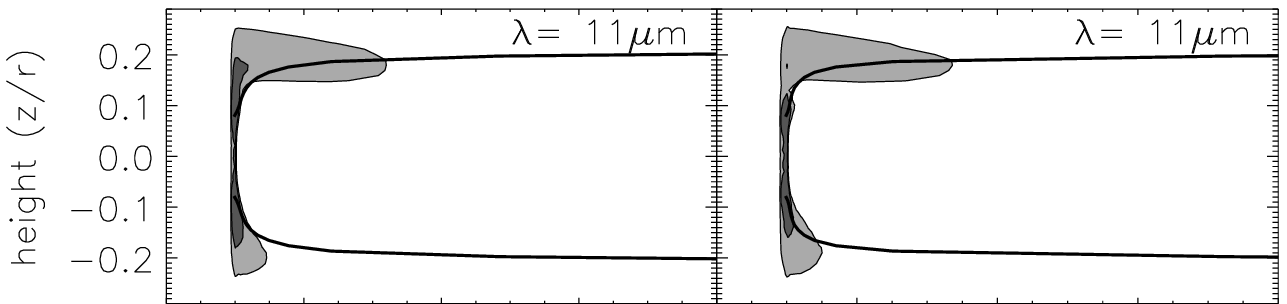}
        \put(-170,61){R20} 
        \put(-60,61){RW30} \\ 
        \includegraphics[width=\linewidth]{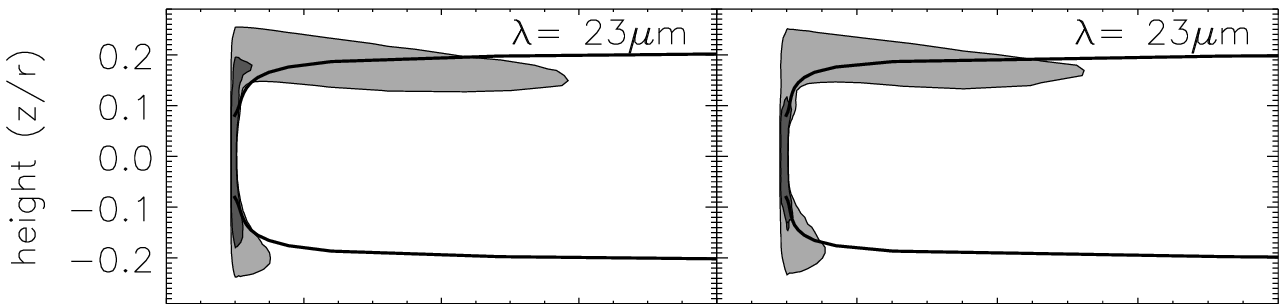}
        \includegraphics[width=\linewidth]{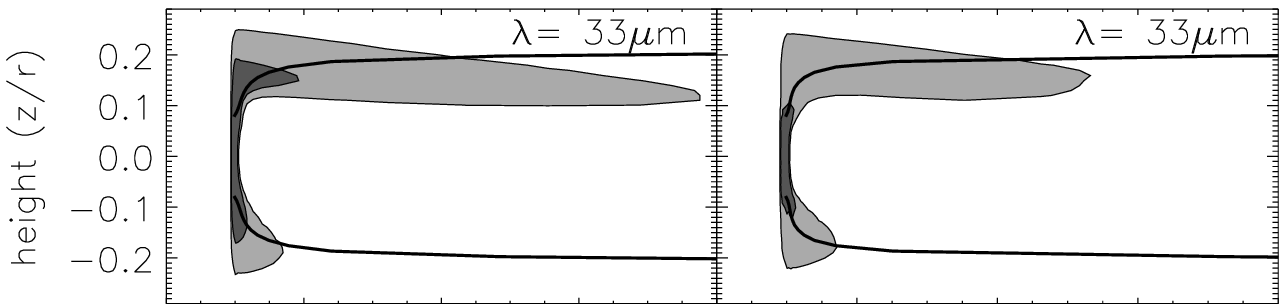}
        \includegraphics[width=\linewidth]{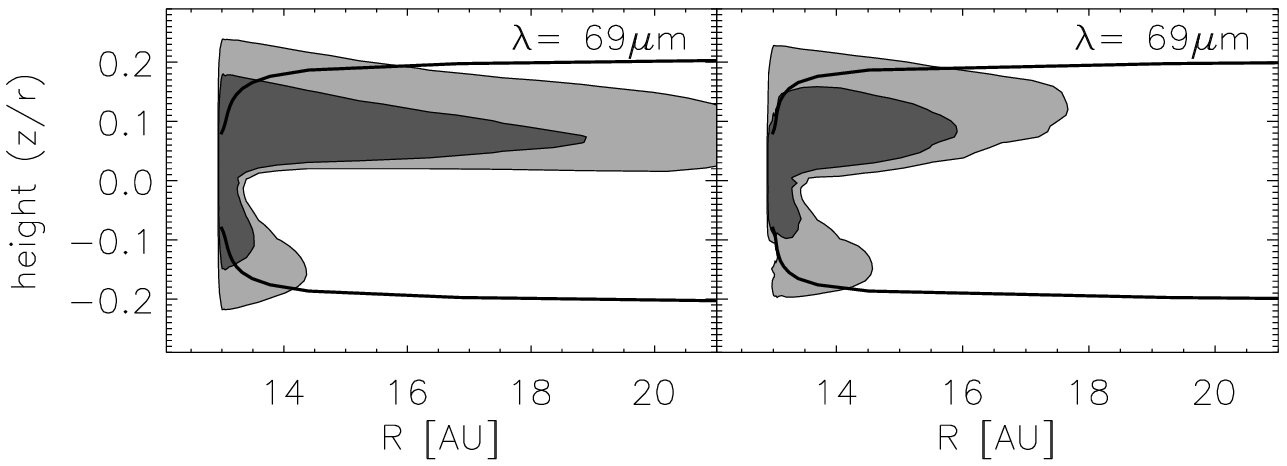}
        \caption[]{Location of emission of the forsterite features at 11, 24, 33 and 69 \um (top to bottom) for models R20 (left) and RW30 (right). Overplotted is the radial $\tau$=1 surface in the optical. Contours are given as a fraction of the maximum emission, and are between 1 and 100 \%. The emission has been integrated over the azimuthal direction. The emission locations are asymmetric around midplane, because the disk is seen from above, and therefore we do not see the bottom half of the disk. Contours are drawn as fraction of the peak intensity at 100\%, 10\% and 1\%, and color coded with dark grey between 10\% and 100\% and light grey between 1\% and 10\%
          \label{fig:vs}}
      \end{figure}

      The contributions to the integrated flux of the features at 11.3, 23.5, 33.5 and 69 \um are plotted in figure \ref{fig:vs} for best-fit models R20 and RW30. All features have a strong contribution from the disk wall, approximately half of the total integrated feature strength at all wavelengths. The remaining emission originates at larger radii at or near the disk surface. Features at longer wavelength can be produced by colder material, and have stronger contributions from larger radii. Due to the lower continuum opacity at longer wavelengths, features also originate from deeper within the disk, especially at larger radii where surface densities become lower.
      
      The shortest wavelength feature at 11.3 \um originates in the warmest regions of the outer disk, mostly in the disk wall. Its surface component extends up to only a few AU outwards of the disk wall, and is  located at the disk optical surface for both models. At 24 and 33 \um, the emission from the disk surface extends farther out, and its contribution to the total features increases up to more than 50\%. As the optical depth of the continuum decreases at these wavelengths, features start originating from deeper in the disk. Both models look similar at 24 \um but start to deviate from each other at 33 \um, as here the cold 100 K forsterite component is not present, truncating most of the emission after 17-18 AU.

      At 69 \um, the optical depth of the continuum is so low that features originate from regions all the way down to the midplane, even in regions just behind the disk wall (14 AU). The total contribution from outside the disk wall is more than 50\% (see also Fig. \ref{fig:radspec}), and peaks towards the midplane with only a very small contribution from the disk optical surface. The model with only warm forsterite doesn't extend as far out because the cold forsterite component in the midplane is not present.

      The detailed analysis of the feature location confirms the simple analysis from the beginning of this section: towards longer wavelengths, the optical depth decreases and features originate from regions deeper in the disk which are hidden from view at shorter wavelengths. Compared to figure \ref{fig:temp}, regions of emission don't reach down as far as the vertical $\tau$=1 surface. The main reason is that disk's inclination is (by definition) not taken into account in calculating the vertical $\tau$=1 surface. When viewing the disk at an inclination angle i, the $\tau$=1 surface of an inclined disk is raised by approximately a factor $\sim$1/cos(i) with respect to a face-on disk.

      \subsubsection{The 69 \um feature}

      \begin{figure}
        \includegraphics[width=\linewidth]{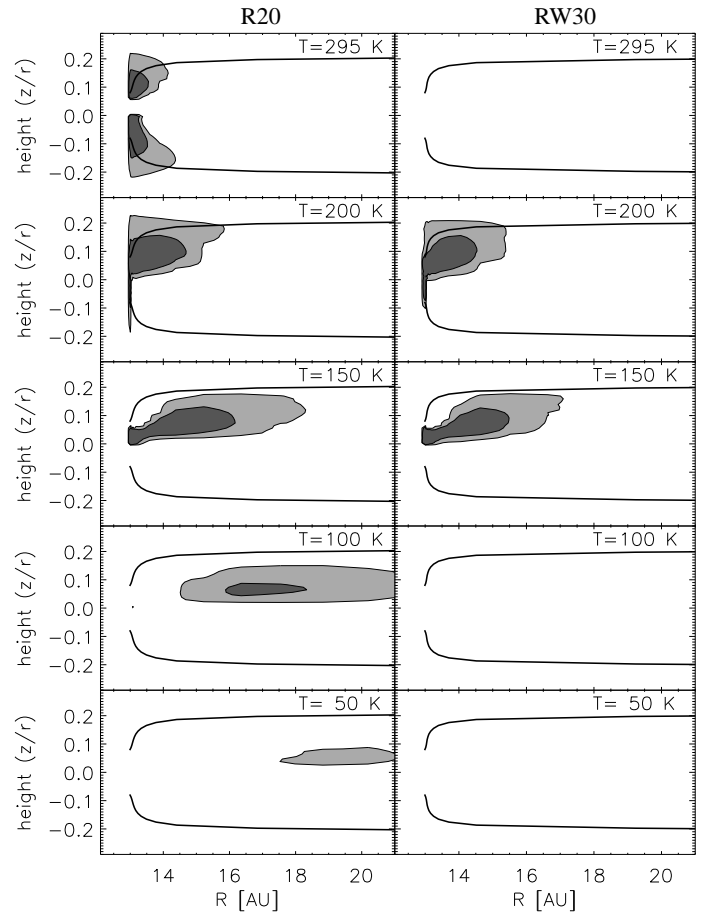}
        \put(-170,335){R20} 
        \put(-60,335){RW30} \\ 
        \caption[]{Location of emission of the forsterite 69 \um features, split into its different temperature components for models R20 (left) and RW30 (right). Overplotted is the radial $\tau$=1 surface in the optical. Contours are given as a fraction of the maximum emission of the total feature, between 1 and 10 \%(light grey) and 10 and 100 \% (dark grey). The emission has been integrated over the azimuthal direction.
          \label{fig:vs69}}
      \end{figure}

      \begin{figure}
        \includegraphics[width=\linewidth]{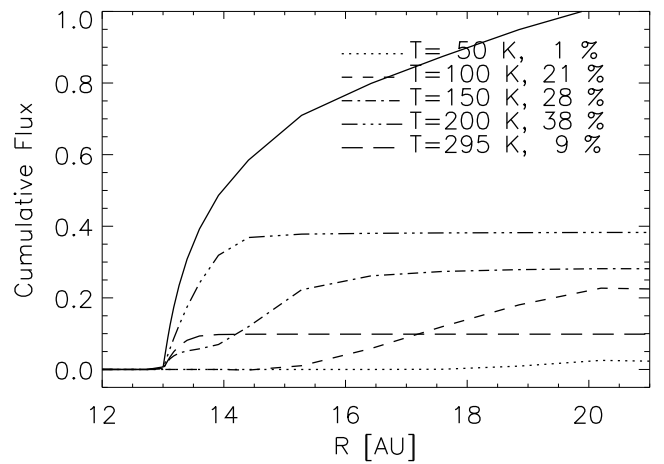}
        \caption[]{Cumulative radial contribution of emission of the forsterite 69 \um feature for model R20. The total flux is scaled to one. The solid line marks the total integrated flux, the dashed and dotted lines mark the contributions of the different temperature components to the integrated flux.
          \label{fig:radspec}}
      \end{figure}

      The location of emission of the different temperature components of the 69 \um feature are displayed in Fig. \ref{fig:vs69} for model R20 and RW30. These are calculated similarly to the shorter wavelength features - continuum subtracting the spectrum in every grid cell - but for each of the five temperature components separately. For model R20, the cumulative radial contribution to the 69 \um feature is plotted in figure \ref{fig:radspec}.

      For model R20, the 295 K component originates from the disk wall (between 13 and 14 AU), except from near the midplane. This is because the temperature in the midplane is slightly lower than at the surface, an effect of the shadow of the inner disk (see Fig. \ref{fig:image}). The 200 K component originates only for a small part in the disk wall. Most of the emission comes from the upper half of the disk between 13 and 15 AU, and peaks below the disk surface. At 150 K, a fraction of the emission comes from the shadowed part of the disk wall, and the bulk from near the midplane below the surface out to 17 AU. The 100 K component does not originate in the disk wall, but from the midplane outwards of 14 AU. The coldest, 50 K component originates from even larger radii($>18 AU$) in the midplane.

      For Model RW30, the 200 K and 150 K components originate in more or less the same regions as for model R20, while the 50 K, 100 K and 295 K components are not present.

      \section{Discussion}\label{sec:discussion}

      \subsection{Forsterite abundance in comparison with previous work}
      The silicate mineralogy of HD 100546 has been addressed in a number of studies (e.g. \citealt{2003A&A...401..577B}, \citealt{2010ApJ...721..431J}), with the underlying assumption that emission features arise in an optically thin surface layer in the disk. Assuming a homogeneous composition for the emitting region, the mass fraction of forsterite in small dust grains is found to be around 5-6\% (\citealt{2005A&A...437..189V}, \citealt{2010ApJ...721..431J}). Because this number can be derived from the mid-infrared spectrum directly by measuring crystalline and amorphous feature strengths, we will refer to this as the \textit{apparent} crystallinity. It is clear that this number is not neccessarily representative for the mass ratio in the entire disk, but an extrapolation is of course tempting. Taking a separate dust composition for the inner and outer disk, \cite{2003A&A...401..577B} find a forsterite abundance in the depleted inner disk of 2\%, compared to  19\% in the outer disk. Since the latter dominates the total mass, the total mass fraction is also 19\%, a factor of 4 higher than the \textit{apparent} crystallinity.

      Our analysis shows that the crystalline forsterite is concentrated towards the disk wall, with locally very high abundances (40-60\%). Although uncertainties in the grain properties and model assumptions can lead to small errors in the derived abundance, the abundances we find are significantly higher than those quoted above. These high abundances are confined to a very narrow region (from 13 AU up to 20-30 AU). Outside of this region - where the bulk of the mass is located - the forsterite abundance is consistent with zero. Therefore, the overall forsterite mass fraction - the mass in crystalline forsterite grains compared to the total disk mass in small dust grains - is around 0.5-0.8\% (See Table \ref{tab:fors_mass}). This is a factor of 10 lower than quoted above.

      \begin{table}
        \centering
        \title{Forsterite mass}

        \begin{tabular}{lll}
          \hline \hline
          Model  & Forsterite mass [M$_{\rm \oplus}$] & Fraction [\%] \\
          \hline
          R20    & 0.27 & 0.8 \\
          RW30   & 0.17 & 0.5 \\
          \hline
          G2     & 0.33 & 1.0 \\
          5\%    & 1.7 & 5.0 \\
          \hline \hline
        \end{tabular}
        \caption{Forsterite mass of our best fit disk models, plus a gradient and constant abundance model for comparison.
          \label{tab:fors_mass}}
      \end{table}

      The reason why such a small amount of forsterite shows up so prominently in the SED is a typical case of window-dressing: The wall reprocesses approximately a third of the total stellar energy captured by the disk, but represents only a few percent of the total disk mass. The wall therefore acts as a display case, showing a strong apparent crystallinity from only a small amount of forsterite with a high local abundance. This may also explain the lack of correlations found in larger samples between \textit{apparent} crystallinity and other disk parameters.

      \subsection{Forsterite abundance in comparison with Solar System comets}
      It has been noted before that the mid-infrared spectrum of comet \object{Hale-Bopp} shows a striking resemblance to that of HD 100546 (e.g. \citealt{1998A&A...332L..25M}). The derived crystallinities are in the range of 25\% to 50\% \citep{1999P&SS...47..773B, 1999ApJ...517.1034W, 2002ApJ...580..579H}. It should be noted that the degree of crystallinity depends on the assumed dust model, see e.g., \cite{hanner}. With the inclusion of larger grains ($>$\um), crystallinities tend to be lower, around 7\% for Hale-Bopp (e.g. \citealt{2005Icar..179..158M, 2003A&A...401..577B}). Because we measure the crystallinity of small dust grains in HD 100546, we can compare the local abundance in the disk wall of 40-60\% with the crystallinity of small grains in comets.

Also other longer period comets with high crystallinities have been observed, for example COMET \object{C/2001 Q4} has a crystallinity of $\sim$30\% \citep{2004ApJ...612L..77W}). The surface of shorter period comets might be subject to amorphization by solar irradiation, but ejecta from the inside comet \object{9P/Tempel} released during the Deep Impact experiment show that these comets can also have a high crystallinity of 20-40\% \citep{2007Icar..191S.432H}.

      Such high crystallinities at these large radii are not expected from radial mixing models \citep{2004A&A...413..571G}, and not observed in typical protoplanetary disks. Because the proposed formation region of comets is beyond the snowline, it coincides with the location of increased crystallinity in the disk wall of HD 100546, making it an ideal place for comet formation.

      \subsection{Crystallinity-gap correlation}
      Both the high abundance and spatial distribution indicate a correlation between the disk geometry and the strong crystalline silicate features in the disk of HD 100546. If the disk gap were filled with amorphous material - so that an illuminated wall is no longer present - the spectral features of forsterite would be weaker by roughly a factor of 3. This might indicate that the strength of crystalline features - the apparent crystallinity - does not reflect the overall crystal abundance, but rather its spatial distribution and the presence of a disk gap.

      Another star which shows a similar geometry and apparent crystallinity is \object{RECX 5} \citep{2010ApJ...723L.243B}. This opens up the possibility that a disk gap-crystallinity correlation also exists for other protoplanetary disks. However, not all disks with gaps show crystalline dust features \citep{2007ApJ...664L.107B}. A reason for this might be that the crystalline features only show up if the gap is located in the right temperature range. A good census of crystallinity and gaps in the right radius/temperature range is needed to establish the presence of such a correlation. The strength of crystalline emission features can be well determined from mid-infrared spectroscopy (e.g. \citealt{2010ApJ...721..431J}). But the characterization of disk gaps from the SED is less trivial, and requires spatially resolved observations \citep{2006ApJ...637L.125V}.

      Aside from the \textit{apparent} crystallinity of HD 100546 - which is one of the highest among Herbig Ae stars (e.g. \citealt{2010ApJ...721..431J}) - also the local abundance of forsterite at tens of AU is unusually high (40-60\% at 13-20 AU). Other Herbig Ae/Be stars show lower forsterite abundances in the outer (2-20 AU) disk, around 10-15\% derived with an optically thin approach \citep{2004Natur.432..479V}, with the exception of \object{HD142527}, which also has a prominent disk gap \citep{2011arXiv1101.5719V}. If all Herbig Ae/Be stars would have the same crystallinity as HD 100546 in the 13-20 AU range - and an equal or higher abundance farther in - they would produce much stronger spectral features than observed. Only if the abundance would increase outwards would it be possible to produce the observed crystallinities in the 0-5\% range with the same high local abundance at 13 AU. However, this is in sharp contrast with observations \citep{2004Natur.432..479V} and radial mixing models (\citealt{2001A&A...378..192G},\citealt{2004A&A...413..571G}), that show a crystallinity that is constant or decreases outwards.

      Whether the spatial distribution of forsterite around HD 100546 is truly unique among Herbig Ae/Be stars requires an analysis of disks with lower crystallinity as well. The Herschel Open Time Key Program DIGIT will observe the 69\um feature in a number of Herbig Ae/Be stars, allowing for a characterization of the spatial distribution of forsterite in these disks in a similar fashion. If there is a mechanism that produces a high crystallinity at 13-20 AU independent of a disk gap, it will be identified in this sample.

      \subsection{Forsterite origin}
      How can such a high abundance of forsterite originate in the disk wall, and is it related to the formation or presence of the disk gap? There are two chemical pathways of producing forsterite: by condensation from the gas phase or by annealing from amorphous olivines. Both require high temperatures above the dust evaporation temperature ($\sim$ 1500 K) or the glass temperature ($\sim$ 1000 K) respectively. These high dust temperatures exist only in the inner regions of protoplanetary disks, but not at 13AU. We can think of two main scenarios, radial transport or in-situ production:

      In the first scenario, forsterite is produced in the inner disk, and transported outwards to 13 AU. The current dust mass in the inner disk is much lower than the total amount of forsterite in the disk wall between 13 and 20 AU. It therefore seems unlikely that radial mixing from the current inner disk can produce such a high crystallinity in the disk wall. As discussed before, it is unlikely that radial mixing has produced a locally high abundance before the opening of the disk gap, as it would have shown up in the mid-infrared spectra. This means that radial transport must have taken place during the formation of the disk gap and depletion of the inner disk. Radiation pressure on the crystalline grains can transport them across the gap and blow them into the disk wall, depleting the inner disk. Also a different mechanism, photophoresis, can transport grains outwards efficiently when the inner disk depletes \citep{2006LPI....37.1558P}. The depleted dust mass from the inner disk (200 times the current inner disk mass, section \ref{sec:SED}) is large enough to explain the crystalline silicates in the disk wall, but it requires extremely efficient radial transport and almost a 100\% crystallinity in the inner 4 AU before the opening of the gap.

      In the second scenario, forsterite is produced locally in the disk wall. Because the equilibrium temperature of the dust in the wall (200 K) is far below the dust evaporation or glass temperature, another mechanism must then be responsible for heating the amorphous dust grains so they can crystallize:
      \begin{itemize}
        \item Flash heating of grains in a stellar outburst \citep{2009Natur.459..224A}. However, HD 100546 shows currently no signs of such outbursts. Such crystals are observed to disappear quickly after formation, which is not the case for HD 100546, where crystals persist over more than a decade of observations.
        \item Shocks in the disk midplane can locally heat the grains above the glass temperature \citep{2002ApJ...565L.109H}. From scattered light images, HD 100546 is known to possess two spiral arms in the outer disk \citep{2001AJ....122.3396G}, which might be related to shocks in the disk midplane that could create forsterite \citep{2002ApJ...565L.109H}. If a companion is responsible for clearing the disk gap, it likely induces a similar spiral wave at the location of the disk wall \citep[e.g. ][]{2003MNRAS.339..993N}, producing forsterite locally.
        \item Amorphous silicates can be crystallized in parent bodies large enough to retain the heat from decay of radioactive elements or accretion (e.g. \citealt{2001M&PS...36..975H}). Micron-size forsterite grains can be created in a collisional cascade of such objects, that can take place in the disk gap or wall (see \citealt{2003A&A...401..577B} for a discussion on this topic.) The amount of forsterite found in the disk wall of HD 100546 requires the destruction of an Earth-like planet or several hundred Pluto-sized objects.
      \end{itemize}

      All plausible scenarios for forsterite formation (shock waves, parent body processing and radial transport during gap formation) are connected to the presence or formation of the disk gap. Therefore we conclude that the anomalously high forsterite abundance in the disk wall of HD 100546 is - one way or the other - connected to the presence of the gap.

      \subsection{Forsterite in the inner disk}\label{sec:disc_inner}
      The mass fraction of forsterite in the inner disk is hard to constrain from our models. The contribution to mid-infrared features in the model is low, and shows a significant contribution only at 11 \um. Regions near the inner rim that are above a temperature of 1000 K should be fully crystalline, but even if the silicates in the entire inner disk are fully crystalline, the contribution is only a few Jy \um at 11 \um. The contribution to longer wavelengths is negligible because the inner disk is too warm, and optical depth effects do not play a role because it is also (vertically) optically thin \citep{2010A&A...511A..75B}. Therefore we cannot put any limits on the crystallinity of the inner disk with the current model and observations.

      \subsection{Thermal contact}\label{sec:tcontact}
      We assume that the different dust species in our model (amorphous silicates, forsterite, carbon) are in thermal contact. However, \cite{2003A&A...401..577B} found with an optically thin approximation that the bulk of the forsterite is at a temperature of \ong 70 K while the other dust species are much warmer, leading them to the conclusion that forsterite cannot be in thermal contact with the other dust species. Because the 2D radiative transfer does not need this cold component due to optical depth effects (see section \ref{sec:opticaldepth}), the forsterite has the same temperature as the other dust species in the disk wall, which means they can be in thermal contact.

      If the forsterite in the outer disk is forced to be out of thermal contact with the other dust species, it does not reach temperatures high enough to explain the shortest wavelength features of forsterite. The reason for this is the very low opacity of forsterite at optical wavelengths compared to the other dust species, especially carbon. Therefore the forsterite is not heated directly by the central star, but only by mid-infrared radiation from the disk surface. It will therefore have a temperature similar to dust species in the midplane (50-100 K), which is too cold to explain any short-wavelength features. Even if the forsterite is moved in to the inner disk, where it would normally be hotter, it falls short of explaining the strength of the short wavelength features (see also section \ref{sec:disc_inner}).

      We can therefore exclude that the dust species are not in thermal contact. \cite{2003A&A...401..577B} argue that the dust in HD100546 is more primitive than in comet \object{Hale-Bopp} because it is out of thermal contact, and that we see the creation of such comets rather than their destruction. Since we exclude that the dust species are not in thermal contact, this implies that the dust in HD 100546 is more similar to comet Hale-Bopp than thought before, and not necessarily more primitive. This opens up again the possibility that we are witnessing the destruction of forsterite-rich comets like Hale-Bopp in HD 100546, rather than their formation.

      \subsection{Iron content}
      Finally,  we use the shift in wavelength required to fit the shape of the 69 \um feature to estimate the iron content of the silicates. For large changes in the iron fraction, the central wavelength shifts linearly with iron content \citep{2003A&A...399.1101K}. The wavelength shifts between 0 and 10\% have not been measured, but relies on the interpolation by \cite{2010A&A...518L.129S}, which assumes the shift in this regime is also linear.

      We also assume that the shape of the 69 \um feature does not change with a small change in iron content. Although this has not been measured in the lab for forsterite with low iron fractions, small changes in iron fraction at higher iron content have been measured \citep{2003A&A...399.1101K}. For example, between 9.3 and 8.6\%, the shape does not change significantly. The feature broadens a little bit between 0 and 8\% iron content, but this change is much smaller than the broadening due to a temperature change (see Fig. \ref{fig:topac}).

      Therefore we expect the shape of the feature not to change significantly for the small changes in iron content ($\textless$0.3\%) found in this paper. Laboratory measurements of forsterite with low iron content are necessary to confirm this.

      \section{Conclusion}\label{sec:conclusion}
      We have studied the spatial distribution of crystalline silicate dust (forsterite) in the gapped disk of Herbig Be star HD 100546. We have used a 2D radiative transfer code to model the dust geometry, and study the effects of optical depth on the forsterite feature strengths in the mid and far-infrared, as well as the shape of the temperature-dependent 69 \um feature observed with Herschel PACS.
      Our conclusions are:
      \begin{enumerate}
        \item 2D radiative transfer effects need to be taken into account when interpreting spectral features over a broad wavelength range. In the case of HD 100546, the bulk of the warm (200-150 K) forsterite that contributes to the 69 \um feature is hidden from view at shorter wavelengths (10-40 \um). An additional cold forsterite component ($\sim$50-70 K) is therefore not necessary to explain the strength of the 69 \um feature.
        \item The forsterite of HD 100546 is located in the disk wall at 13 AU, with extremely high local abundance. Based on mid-infrared features strengths and the shape of the 69 \um feature, we arrive at two best-fit models:
         \\a) a model with 40\% forsterite between 13 and 20 AU.
         \\b) a model with 60\% forsterite between 13 and 30 AU but restricted to regions where the temperatures are around 150 and 200 K. This excludes the cold midplane outside of the disk wall.
        \item All forsterite features show a strong contribution from the disk wall, but also probe a region farther out. The 69 \um feature from this region originates from \textit{below} the disk surface. The shorter wavelength features from this region originate \textit{at} the disk surface.
        \item The mass fraction of forsterite compared to the total disk mass is low - around 0.5-0.8\% - a factor of 10 lower than the \textit{apparent} crystallinity derived from and optically thin analysis of relative feature strengths in the mid-infrared. The wall at 13 AU acts as a display case for showing a high local forsterite abundance.
        \item The crystalline forsterite in the disk of HD 100546 is in thermal contact with the other dust species.
        \item We cannot rule out forsterite that is completely iron free. Our best fits indicate an iron content of $\textless$0.3 \% of the crystalline olivines.
        \item Due to a degeneracy in the modelling between iron content and dust temperature, it is not straightforward to use the 69 \um feature as an independent temperature indicator. 
        \item A disk model in hydrostatic equilibrium (with T$_{\rm gas}$=T$_{\rm dust}$) can self-consistently explain the SED and mid-infrared imaging. Our best fit has a surface density power law $\propto r^{-1}$ and an inner disk depletion factor of 200. Additionally, continuum opacity sources - such as carbon - must have a low abundance to fit the low flux at 8 \um.
        \item  The \textit{apparent} crystallinity of dust as measured from the ratio of crystalline and amorphous components of the mid-infrared features is not necessarily a good measure of the \textit{real} crystallinity, the mass fraction of small dust grains in the disk that is crystalline. Both disk geometry and spatial distribution of the crystals can significantly distort the picture, and a one-to-one correlation between crystal feature strength and disk state is not to be expected. However, in our case, the detailed analysis shows a highly localized region with high abundance, which may very well inform us about special processing in this region, possibly linked to planet formation.
      \end{enumerate}

      \begin{acknowledgements}
        This research project is financially supported by a joint grant from the
        Netherlands Research School for Astronomy (NOVA) and the Netherlands
        Institute for Space Research (SRON). 
        J.C.A. acknowledges CNES and PNPS for financial support
      \end{acknowledgements}
      %
      %
      \bibliographystyle{aa} 
      \bibliography{16770}

\begin{thebibliography}{81}
\expandafter\ifx\csname natexlab\endcsname\relax\def\natexlab#1{#1}\fi

\bibitem[{{{\'A}brah{\'a}m} {et~al.}(2009){{\'A}brah{\'a}m}, {Juh{\'a}sz},
  {Dullemond}, {K{\'o}sp{\'a}l}, {van Boekel}, {Bouwman}, {Henning},
  {Mo{\'o}r}, {Mosoni}, {Sicilia-Aguilar}, \& {Sipos}}]{2009Natur.459..224A}
{{\'A}brah{\'a}m}, P., {Juh{\'a}sz}, A., {Dullemond}, C.~P., {et~al.} 2009,
  \nat, 459, 224

\bibitem[{{Acke} {et~al.}(2009){Acke}, {Min}, {van den Ancker}, {Bouwman},
  {Ochsendorf}, {Juhasz}, \& {Waters}}]{2009A&A...502L..17A}
{Acke}, B., {Min}, M., {van den Ancker}, M.~E., {et~al.} 2009, \aap, 502, L17

\bibitem[{{Acke} \& {van den Ancker}(2006)}]{2006A&A...449..267A}
{Acke}, B. \& {van den Ancker}, M.~E. 2006, \aap, 449, 267

\bibitem[{{Ardila} {et~al.}(2007){Ardila}, {Golimowski}, {Krist}, {Clampin},
  {Ford}, \& {Illingworth}}]{2007ApJ...665..512A}
{Ardila}, D.~R., {Golimowski}, D.~A., {Krist}, J.~E., {et~al.} 2007, \apj, 665,
  512

\bibitem[{{Augereau} {et~al.}(2001){Augereau}, {Lagrange}, {Mouillet}, \&
  {M{\'e}nard}}]{2001A&A...365...78A}
{Augereau}, J.~C., {Lagrange}, A.~M., {Mouillet}, D., \& {M{\'e}nard}, F. 2001,
  \aap, 365, 78

\bibitem[{{Benisty} {et~al.}(2010){Benisty}, {Tatulli}, {M{\'e}nard}, \&
  {Swain}}]{2010A&A...511A..75B}
{Benisty}, M., {Tatulli}, E., {M{\'e}nard}, F., \& {Swain}, M.~R. 2010, \aap,
  511, A75+

\bibitem[{{Bjorkman} \& {Wood}(2001)}]{2001ApJ...554..615B}
{Bjorkman}, J.~E. \& {Wood}, K. 2001, \apj, 554, 615

\bibitem[{{Bouwman} {et~al.}(2003){Bouwman}, {de Koter}, {Dominik}, \&
  {Waters}}]{2003A&A...401..577B}
{Bouwman}, J., {de Koter}, A., {Dominik}, C., \& {Waters}, L.~B.~F.~M. 2003,
  \aap, 401, 577

\bibitem[{{Bouwman} {et~al.}(2010){Bouwman}, {Lawson}, {Juh{\'a}sz}, {Dominik},
  {Feigelson}, {Henning}, {Tielens}, \& {Waters}}]{2010ApJ...723L.243B}
{Bouwman}, J., {Lawson}, W.~A., {Juh{\'a}sz}, A., {et~al.} 2010, \apjl, 723,
  L243

\bibitem[{{Bouwman} {et~al.}(2001){Bouwman}, {Meeus}, {de Koter}, {Hony},
  {Dominik}, \& {Waters}}]{2001A&A...375..950B}
{Bouwman}, J., {Meeus}, G., {de Koter}, A., {et~al.} 2001, \aap, 375, 950

\bibitem[{{Bowey} {et~al.}(2002){Bowey}, {Barlow}, {Molster}, {Hofmeister},
  {Lee}, {Tucker}, {Lim}, {Ade}, \& {Waters}}]{2002MNRAS.331L...1B}
{Bowey}, J.~E., {Barlow}, M.~J., {Molster}, F.~J., {et~al.} 2002, \mnras, 331,
  L1

\bibitem[{{Brittain} {et~al.}(2009){Brittain}, {Najita}, \&
  {Carr}}]{2009ApJ...702...85B}
{Brittain}, S.~D., {Najita}, J.~R., \& {Carr}, J.~S. 2009, \apj, 702, 85

\bibitem[{{Brown} {et~al.}(2007){Brown}, {Blake}, {Dullemond}, {Mer{\'{\i}}n},
  {Augereau}, {Boogert}, {Evans}, {Geers}, {Lahuis}, {Kessler-Silacci},
  {Pontoppidan}, \& {van Dishoeck}}]{2007ApJ...664L.107B}
{Brown}, J.~M., {Blake}, G.~A., {Dullemond}, C.~P., {et~al.} 2007, \apjl, 664,
  L107

\bibitem[{{Brucato} {et~al.}(1999){Brucato}, {Colangeli}, {Mennella},
  {Palumbo}, \& {Bussoletti}}]{1999P&SS...47..773B}
{Brucato}, J.~R., {Colangeli}, L., {Mennella}, V., {Palumbo}, P., \&
  {Bussoletti}, E. 1999, \planss, 47, 773

\bibitem[{{Crovisier} {et~al.}(1997){Crovisier}, {Leech}, {Bockelee-Morvan},
  {Brooke}, {Hanner}, {Altieri}, {Keller}, \& {Lellouch}}]{1997Sci...275.1904C}
{Crovisier}, J., {Leech}, K., {Bockelee-Morvan}, D., {et~al.} 1997, Science,
  275, 1904

\bibitem[{{Dominik} {et~al.}(2003){Dominik}, {Dullemond}, {Waters}, \&
  {Walch}}]{2003A&A...398..607D}
{Dominik}, C., {Dullemond}, C.~P., {Waters}, L.~B.~F.~M., \& {Walch}, S. 2003,
  \aap, 398, 607

\bibitem[{{Dorschner} {et~al.}(1995){Dorschner}, {Begemann}, {Henning},
  {Jaeger}, \& {Mutschke}}]{1995A&A...300..503D}
{Dorschner}, J., {Begemann}, B., {Henning}, T., {Jaeger}, C., \& {Mutschke}, H.
  1995, \aap, 300, 503

\bibitem[{{Espaillat} {et~al.}(2010){Espaillat}, {D'Alessio}, {Hern{\'a}ndez},
  {Nagel}, {Luhman}, {Watson}, {Calvet}, {Muzerolle}, \&
  {McClure}}]{2010ApJ...717..441E}
{Espaillat}, C., {D'Alessio}, P., {Hern{\'a}ndez}, J., {et~al.} 2010, \apj,
  717, 441

\bibitem[{{Fabian} {et~al.}(2000){Fabian}, {J{\"a}ger}, {Henning}, {Dorschner},
  \& {Mutschke}}]{2000A&A...364..282F}
{Fabian}, D., {J{\"a}ger}, C., {Henning}, T., {Dorschner}, J., \& {Mutschke},
  H. 2000, \aap, 364, 282

\bibitem[{{Gail}(2001)}]{2001A&A...378..192G}
{Gail}, H. 2001, \aap, 378, 192

\bibitem[{{Gail}(2004)}]{2004A&A...413..571G}
{Gail}, H. 2004, \aap, 413, 571

\bibitem[{{Glauser} {et~al.}(2009){Glauser}, {G{\"u}del}, {Watson}, {Henning},
  {Schegerer}, {Wolf}, {Audard}, \& {Baldovin-Saavedra}}]{2009A&A...508..247G}
{Glauser}, A.~M., {G{\"u}del}, M., {Watson}, D.~M., {et~al.} 2009, \aap, 508,
  247

\bibitem[{{Grady} {et~al.}(2001){Grady}, {Polomski}, {Henning}, {Stecklum},
  {Woodgate}, {Telesco}, {Pi{\~n}a}, {Gull}, {Boggess}, {Bowers}, {Bruhweiler},
  {Clampin}, {Danks}, {Green}, {Heap}, {Hutchings}, {Jenkins}, {Joseph},
  {Kaiser}, {Kimble}, {Kraemer}, {Lindler}, {Linsky}, {Maran}, {Moos}, {Plait},
  {Roesler}, {Timothy}, \& {Weistrop}}]{2001AJ....122.3396G}
{Grady}, C.~A., {Polomski}, E.~F., {Henning}, T., {et~al.} 2001, \aj, 122, 3396

\bibitem[{{Grady} {et~al.}(2005){Grady}, {Woodgate}, {Heap}, {Bowers}, {Nuth},
  {Herczeg}, \& {Hill}}]{2005ApJ...620..470G}
{Grady}, C.~A., {Woodgate}, B., {Heap}, S.~R., {et~al.} 2005, \apj, 620, 470

\bibitem[{{Hanner} \& {Zolensky}(2010)}]{hanner}
{Hanner}, M.~S. \& {Zolensky}, M.~E. 2010, in Lecture Notes in Physics, Berlin
  Springer Verlag, Vol. 609, Astromineralogy, ed. {T.~K.~Henning}

\bibitem[{{Harker} \& {Desch}(2002)}]{2002ApJ...565L.109H}
{Harker}, D.~E. \& {Desch}, S.~J. 2002, \apjl, 565, L109

\bibitem[{{Harker} {et~al.}(2002){Harker}, {Wooden}, {Woodward}, \&
  {Lisse}}]{2002ApJ...580..579H}
{Harker}, D.~E., {Wooden}, D.~H., {Woodward}, C.~E., \& {Lisse}, C.~M. 2002,
  \apj, 580, 579

\bibitem[{{Harker} {et~al.}(2007){Harker}, {Woodward}, {Wooden}, {Fisher}, \&
  {Trujillo}}]{2007Icar..191S.432H}
{Harker}, D.~E., {Woodward}, C.~E., {Wooden}, D.~H., {Fisher}, R.~S., \&
  {Trujillo}, C.~A. 2007, \icarus, 191, 432

\bibitem[{{Henning} {et~al.}(1998){Henning}, {Burkert}, {Launhardt}, {Leinert},
  \& {Stecklum}}]{1998A&A...336..565H}
{Henning}, T., {Burkert}, A., {Launhardt}, R., {Leinert}, C., \& {Stecklum}, B.
  1998, \aap, 336, 565

\bibitem[{{Henning} \& {Stognienko}(1996)}]{1996A&A...311..291H}
{Henning}, T. \& {Stognienko}, R. 1996, \aap, 311, 291

\bibitem[{{Hughes} {et~al.}(2008){Hughes}, {Wilner}, {Qi}, \&
  {Hogerheijde}}]{2008ApJ...678.1119H}
{Hughes}, A.~M., {Wilner}, D.~J., {Qi}, C., \& {Hogerheijde}, M.~R. 2008, \apj,
  678, 1119

\bibitem[{{Huss} {et~al.}(2001){Huss}, {MacPherson}, {Wasserburg}, {Russell},
  \& {Srinivasan}}]{2001M&PS...36..975H}
{Huss}, G.~R., {MacPherson}, G.~J., {Wasserburg}, G.~J., {Russell}, S.~S., \&
  {Srinivasan}, G. 2001, Meteoritics and Planetary Science, 36, 975

\bibitem[{{Juh{\'a}sz} {et~al.}(2010){Juh{\'a}sz}, {Bouwman}, {Henning},
  {Acke}, {van den Ancker}, {Meeus}, {Dominik}, {Min}, {Tielens}, \&
  {Waters}}]{2010ApJ...721..431J}
{Juh{\'a}sz}, A., {Bouwman}, J., {Henning}, T., {et~al.} 2010, \apj, 721, 431

\bibitem[{{Kalas} {et~al.}(2008){Kalas}, {Graham}, {Chiang}, {Fitzgerald},
  {Clampin}, {Kite}, {Stapelfeldt}, {Marois}, \& {Krist}}]{2008Sci...322.1345K}
{Kalas}, P., {Graham}, J.~R., {Chiang}, E., {et~al.} 2008, Science, 322, 1345

\bibitem[{{Kemper} {et~al.}(2004){Kemper}, {Vriend}, \&
  {Tielens}}]{2004ApJ...609..826K}
{Kemper}, F., {Vriend}, W.~J., \& {Tielens}, A.~G.~G.~M. 2004, \apj, 609, 826

\bibitem[{{Koike} {et~al.}(2003){Koike}, {Chihara}, {Tsuchiyama}, {Suto},
  {Sogawa}, \& {Okuda}}]{2003A&A...399.1101K}
{Koike}, C., {Chihara}, H., {Tsuchiyama}, A., {et~al.} 2003, \aap, 399, 1101

\bibitem[{{Koike} {et~al.}(2006){Koike}, {Mutschke}, {Suto}, {Naoi}, {Chihara},
  {Henning}, {J{\"a}ger}, {Tsuchiyama}, {Dorschner}, \&
  {Okuda}}]{2006A&A...449..583K}
{Koike}, C., {Mutschke}, H., {Suto}, H., {et~al.} 2006, \aap, 449, 583

\bibitem[{{Lagage} {et~al.}(2004){Lagage}, {Pel}, {Authier}, {Belorgey},
  {Claret}, {Doucet}, {Dubreuil}, {Durand}, {Elswijk}, {Girardot}, {K{\"a}ufl},
  {Kroes}, {Lortholary}, {Lussignol}, {Marchesi}, {Pantin}, {Peletier},
  {Pirard}, {Pragt}, {Rio}, {Schoenmaker}, {Siebenmorgen}, {Silber}, {Smette},
  {Sterzik}, \& {Veyssiere}}]{2004Msngr.117...12L}
{Lagage}, P.~O., {Pel}, J.~W., {Authier}, M., {et~al.} 2004, The Messenger,
  117, 12

\bibitem[{{Lagrange} {et~al.}(2010){Lagrange}, {Bonnefoy}, {Chauvin}, {Apai},
  {Ehrenreich}, {Boccaletti}, {Gratadour}, {Rouan}, {Mouillet}, {Lacour}, \&
  {Kasper}}]{2010Sci...329...57L}
{Lagrange}, A., {Bonnefoy}, M., {Chauvin}, G., {et~al.} 2010, Science, 329, 57

\bibitem[{{Lee} {et~al.}(2010){Lee}, {Bergin}, \&
  {Nomura}}]{2010ApJ...710L..21L}
{Lee}, J., {Bergin}, E.~A., \& {Nomura}, H. 2010, \apjl, 710, L21

\bibitem[{{Liu} {et~al.}(2003){Liu}, {Hinz}, {Meyer}, {Mamajek}, {Hoffmann}, \&
  {Hora}}]{2003ApJ...598L.111L}
{Liu}, W.~M., {Hinz}, P.~M., {Meyer}, M.~R., {et~al.} 2003, \apjl, 598, L111

\bibitem[{{Lucy}(1999)}]{1999A&A...345..211L}
{Lucy}, L.~B. 1999, \aap, 345, 211

\bibitem[{{Malfait} {et~al.}(1998){Malfait}, {Waelkens}, {Waters},
  {Vandenbussche}, {Huygen}, \& {de Graauw}}]{1998A&A...332L..25M}
{Malfait}, K., {Waelkens}, C., {Waters}, L.~B.~F.~M., {et~al.} 1998, \aap, 332,
  L25

\bibitem[{{Marois} {et~al.}(2008){Marois}, {Macintosh}, {Barman}, {Zuckerman},
  {Song}, {Patience}, {Lafreni{\`e}re}, \& {Doyon}}]{2008Sci...322.1348M}
{Marois}, C., {Macintosh}, B., {Barman}, T., {et~al.} 2008, Science, 322, 1348

\bibitem[{{Meijer} {et~al.}(2008){Meijer}, {Dominik}, {de Koter}, {Dullemond},
  {van Boekel}, \& {Waters}}]{2008A&A...492..451M}
{Meijer}, J., {Dominik}, C., {de Koter}, A., {et~al.} 2008, \aap, 492, 451

\bibitem[{{Min} {et~al.}(2009){Min}, {Dullemond}, {Dominik}, {de Koter}, \&
  {Hovenier}}]{2009A&A...497..155M}
{Min}, M., {Dullemond}, C.~P., {Dominik}, C., {de Koter}, A., \& {Hovenier},
  J.~W. 2009, \aap, 497, 155

\bibitem[{{Min} {et~al.}(2005{\natexlab{a}}){Min}, {Hovenier}, \& {de
  Koter}}]{2005A&A...432..909M}
{Min}, M., {Hovenier}, J.~W., \& {de Koter}, A. 2005{\natexlab{a}}, \aap, 432,
  909

\bibitem[{{Min} {et~al.}(2005{\natexlab{b}}){Min}, {Hovenier}, {de Koter},
  {Waters}, \& {Dominik}}]{2005Icar..179..158M}
{Min}, M., {Hovenier}, J.~W., {de Koter}, A., {Waters}, L.~B.~F.~M., \&
  {Dominik}, C. 2005{\natexlab{b}}, \icarus, 179, 158

\bibitem[{{Min} {et~al.}(2007){Min}, {Waters}, {de Koter}, {Hovenier},
  {Keller}, \& {Markwick-Kemper}}]{2007A&A...462..667M}
{Min}, M., {Waters}, L.~B.~F.~M., {de Koter}, A., {et~al.} 2007, \aap, 462, 667

\bibitem[{{Morlok} {et~al.}(2010){Morlok}, {Koike}, {Tomioka}, {Mann}, \&
  {Tomeoka}}]{2010Icar..207...45M}
{Morlok}, A., {Koike}, C., {Tomioka}, N., {Mann}, I., \& {Tomeoka}, K. 2010,
  \icarus, 207, 45

\bibitem[{{Mulders} {et~al.}(2010){Mulders}, {Dominik}, \&
  {Min}}]{2010A&A...512A..11M}
{Mulders}, G.~D., {Dominik}, C., \& {Min}, M. 2010, \aap, 512, A11+

\bibitem[{{Mutschke} {et~al.}(1998){Mutschke}, {Begemann}, {Dorschner},
  {Guertler}, {Gustafson}, {Henning}, \& {Stognienko}}]{1998A&A...333..188M}
{Mutschke}, H., {Begemann}, B., {Dorschner}, J., {et~al.} 1998, \aap, 333, 188

\bibitem[{{Nelson} \& {Papaloizou}(2003)}]{2003MNRAS.339..993N}
{Nelson}, R.~P. \& {Papaloizou}, J.~C.~B. 2003, \mnras, 339, 993

\bibitem[{{Oliveira} {et~al.}(2010){Oliveira}, {Pontoppidan}, {Mer{\'{\i}}n},
  {van Dishoeck}, {Lahuis}, {Geers}, {J{\o}rgensen}, {Olofsson}, {Augereau}, \&
  {Brown}}]{2010ApJ...714..778O}
{Oliveira}, I., {Pontoppidan}, K.~M., {Mer{\'{\i}}n}, B., {et~al.} 2010, \apj,
  714, 778

\bibitem[{{Olofsson} {et~al.}(2010){Olofsson}, {Augereau}, {van Dishoeck},
  {Mer{\'{\i}}n}, {Grosso}, {M{\'e}nard}, {Blake}, \&
  {Monin}}]{2010A&A...520A..39O}
{Olofsson}, J., {Augereau}, J., {van Dishoeck}, E.~F., {et~al.} 2010, \aap,
  520, A39+

\bibitem[{{Olofsson} {et~al.}(2009){Olofsson}, {Augereau}, {van Dishoeck},
  {Mer{\'{\i}}n}, {Lahuis}, {Kessler-Silacci}, {Dullemond}, {Oliveira},
  {Blake}, {Boogert}, {Brown}, {Evans}, {Geers}, {Knez}, {Monin}, \&
  {Pontoppidan}}]{2009A&A...507..327O}
{Olofsson}, J., {Augereau}, J., {van Dishoeck}, E.~F., {et~al.} 2009, \aap,
  507, 327

\bibitem[{{Pani{\'c}} {et~al.}(2010){Pani{\'c}}, {van Dishoeck}, {Hogerheijde},
  {Belloche}, {G{\"u}sten}, {Boland}, \& {Baryshev}}]{2010A&A...519A.110P}
{Pani{\'c}}, O., {van Dishoeck}, E.~F., {Hogerheijde}, M.~R., {et~al.} 2010,
  \aap, 519, A110+

\bibitem[{{Pantin}(2010)}]{Pantin2010}
{Pantin}, E. 2010, to be published as Habilitation $\grave{\rm a}$ Diriger des
  Recherches (HDR) manuscript, Paris VII-Diderot University,
  http://tel.archives-ouvertes.fr

\bibitem[{{Pantin} {et~al.}(2008){Pantin}, {Doucet}, {K{\"a}ufl}, {Lagage},
  {Siebenmorgen}, \& {Sterzik}}]{2008SPIE.7014E..70P}
{Pantin}, E., {Doucet}, C., {K{\"a}ufl}, H.~U., {et~al.} 2008, in Society of
  Photo-Optical Instrumentation Engineers (SPIE) Conference Series, Vol. 7014,
  Society of Photo-Optical Instrumentation Engineers (SPIE) Conference Series

\bibitem[{{Pantin} {et~al.}(2009){Pantin}, {Siebenmorgen}, {K{\"a}ufl}, \&
  {Sterzik}}]{2009svlt.conf..261P}
{Pantin}, E., {Siebenmorgen}, R., {K{\"a}ufl}, H.~U., \& {Sterzik}, M. 2009, in
  Science with the VLT in the ELT Era, ed. A.~{Moorwood}, 261--+

\bibitem[{{Pantin} {et~al.}(2000){Pantin}, {Waelkens}, \&
  {Lagage}}]{2000A&A...361L...9P}
{Pantin}, E., {Waelkens}, C., \& {Lagage}, P.~O. 2000, \aap, 361, L9

\bibitem[{{Petit} {et~al.}(2006){Petit}, {Mousis}, {Alibert}, \&
  {Horner}}]{2006LPI....37.1558P}
{Petit}, J., {Mousis}, O., {Alibert}, Y., \& {Horner}, J. 2006, in Lunar and
  Planetary Institute Science Conference Abstracts, Vol.~37, 37th Annual Lunar
  and Planetary Science Conference, ed. {S.~Mackwell \& E.~Stansbery}, 1558--+

\bibitem[{{Pinte} {et~al.}(2009){Pinte}, {Harries}, {Min}, {Watson},
  {Dullemond}, {Woitke}, {M{\'e}nard}, \&
  {Dur{\'a}n-Rojas}}]{2009A&A...498..967P}
{Pinte}, C., {Harries}, T.~J., {Min}, M., {et~al.} 2009, \aap, 498, 967

\bibitem[{{Preibisch} {et~al.}(1993){Preibisch}, {Ossenkopf}, {Yorke}, \&
  {Henning}}]{1993A&A...279..577P}
{Preibisch}, T., {Ossenkopf}, V., {Yorke}, H.~W., \& {Henning}, T. 1993, \aap,
  279, 577

\bibitem[{{Servoin} \& {Piriou}(1973)}]{1973PSSBR..55..677S}
{Servoin}, J.~L. \& {Piriou}, B. 1973, Physica Status Solidi B Basic Research,
  55, 677

\bibitem[{{Stevenson}(1990)}]{1990ApJ...348..730S}
{Stevenson}, D.~J. 1990, \apj, 348, 730

\bibitem[{{Sturm} {et~al.}(2010){Sturm}, {Bouwman}, {Henning}, {Evans}, {Acke},
  {Mulders}, {Waters}, {van Dishoeck}, {Meeus}, {Green}, {Augereau},
  {Olofsson}, {Salyk}, {Najita}, {Herczeg}, {van Kempen}, {Kristensen},
  {Dominik}, {Carr}, {Waelkens}, {Bergin}, {Blake}, {Brown}, {Chen}, {Cieza},
  {Dunham}, {Glassgold}, {G{\"u}del}, {Harvey}, {Hogerheijde}, {Jaffe},
  {J{\o}rgensen}, {Kim}, {Knez}, {Lacy}, {Lee}, {Maret}, {Meijerink},
  {Mer{\'{\i}}n}, {Mundy}, {Pontoppidan}, {Visser}, \&
  {Y{\i}ld{\i}z}}]{2010A&A...518L.129S}
{Sturm}, B., {Bouwman}, J., {Henning}, T., {et~al.} 2010, \aap, 518, L129+

\bibitem[{{Suto} {et~al.}(2006){Suto}, {Sogawa}, {Tachibana}, {Koike},
  {Karoji}, {Tsuchiyama}, {Chihara}, {Mizutani}, {Akedo}, {Ogiso}, {Fukui}, \&
  {Ohara}}]{2006MNRAS.370.1599S}
{Suto}, H., {Sogawa}, H., {Tachibana}, S., {et~al.} 2006, \mnras, 370, 1599

\bibitem[{{Tatulli} {et~al.}(2011){Tatulli}, {Benisty}, {M{\'e}nard},
  {Varni{\`e}re}, {Martin-Zaidi}, {Thi}, {Pinte}, {Massi}, {Weigelt},
  {Hofmann}, \& {Petrov}}]{2011arXiv1104.0905T}
{Tatulli}, E., {Benisty}, M., {M{\'e}nard}, F., {et~al.} 2011, ArXiv e-prints

\bibitem[{{van Boekel} {et~al.}(2004){van Boekel}, {Min}, {Leinert}, {Waters},
  {Richichi}, {Chesneau}, {Dominik}, {Jaffe}, {Dutrey}, {Graser}, {Henning},
  {de Jong}, {K{\"o}hler}, {de Koter}, {Lopez}, {Malbet}, {Morel}, {Paresce},
  {Perrin}, {Preibisch}, {Przygodda}, {Sch{\"o}ller}, \&
  {Wittkowski}}]{2004Natur.432..479V}
{van Boekel}, R., {Min}, M., {Leinert}, C., {et~al.} 2004, \nat, 432, 479

\bibitem[{{van Boekel} {et~al.}(2005){van Boekel}, {Min}, {Waters}, {de Koter},
  {Dominik}, {van den Ancker}, \& {Bouwman}}]{2005A&A...437..189V}
{van Boekel}, R., {Min}, M., {Waters}, L.~B.~F.~M., {et~al.} 2005, \aap, 437,
  189

\bibitem[{{van den Ancker} {et~al.}(1997){van den Ancker}, {The}, {Tjin A
  Djie}, {Catala}, {de Winter}, {Blondel}, \& {Waters}}]{1997A&A...324L..33V}
{van den Ancker}, M.~E., {The}, P.~S., {Tjin A Djie}, H.~R.~E., {et~al.} 1997,
  \aap, 324, L33

\bibitem[{{van der Plas} {et~al.}(2009){van der Plas}, {van den Ancker},
  {Acke}, {Carmona}, {Dominik}, {Fedele}, \& {Waters}}]{2009A&A...500.1137V}
{van der Plas}, G., {van den Ancker}, M.~E., {Acke}, B., {et~al.} 2009, \aap,
  500, 1137

\bibitem[{{Varni{\`e}re} {et~al.}(2006){Varni{\`e}re}, {Bjorkman}, {Frank},
  {Quillen}, {Carciofi}, {Whitney}, \& {Wood}}]{2006ApJ...637L.125V}
{Varni{\`e}re}, P., {Bjorkman}, J.~E., {Frank}, A., {et~al.} 2006, \apjl, 637,
  L125

\bibitem[{{Verhoeff} {et~al.}(2010){Verhoeff}, {Min}, {Acke}, {van Boekel},
  {Pantin}, {Waters}, {Tielens}, {van den Ancker}, {Mulders}, {de Koter}, \&
  {Bouwman}}]{2010A&A...516A..48V}
{Verhoeff}, A.~P., {Min}, M., {Acke}, B., {et~al.} 2010, \aap, 516, A48+

\bibitem[{{Verhoeff} {et~al.}(2011){Verhoeff}, {Min}, {Pantin}, {Waters},
  {Tielens}, {Honda}, {Fujiwara}, {Bouwman}, {van Boekel}, {Dougherty}, {de
  Koter}, {Dominik}, \& {Mulders}}]{2011arXiv1101.5719V}
{Verhoeff}, A.~P., {Min}, M., {Pantin}, E., {et~al.} 2011, ArXiv e-prints

\bibitem[{{Vinkovi{\'c}} {et~al.}(2006){Vinkovi{\'c}}, {Ivezi{\'c}},
  {Jurki{\'c}}, \& {Elitzur}}]{2006ApJ...636..348V}
{Vinkovi{\'c}}, D., {Ivezi{\'c}}, {\v Z}., {Jurki{\'c}}, T., \& {Elitzur}, M.
  2006, \apj, 636, 348

\bibitem[{{Waters} \& {Waelkens}(1998)}]{1998ARA&A..36..233W}
{Waters}, L.~B.~F.~M. \& {Waelkens}, C. 1998, \araa, 36, 233

\bibitem[{{Watson}(2009)}]{2009ASPC..414...77W}
{Watson}, D. 2009, in Astronomical Society of the Pacific Conference Series,
  Vol. 414, Astronomical Society of the Pacific Conference Series, ed.
  {T.~Henning, E.~Gr{\"u}n, \& J.~Steinacker}, 77--+

\bibitem[{{Wooden} {et~al.}(1999){Wooden}, {Harker}, {Woodward}, {Butner},
  {Koike}, {Witteborn}, \& {McMurtry}}]{1999ApJ...517.1034W}
{Wooden}, D.~H., {Harker}, D.~E., {Woodward}, C.~E., {et~al.} 1999, \apj, 517,
  1034

\bibitem[{{Wooden} {et~al.}(2004){Wooden}, {Woodward}, \&
  {Harker}}]{2004ApJ...612L..77W}
{Wooden}, D.~H., {Woodward}, C.~E., \& {Harker}, D.~E. 2004, \apjl, 612, L77

\end{thebibliography}
      %
      %
      \begin{appendix}

        \section{VISIR imaging}\label{sec:visir}
        \subsection{VISIR observations}\label{sec:visir_obs}
\begin{figure}

        \centering \includegraphics[width=0.6\linewidth]{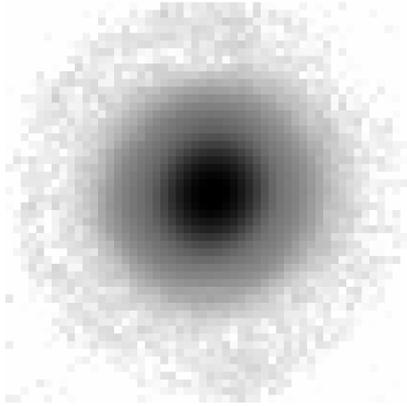}
        \caption[]{VISIR Q band image at 18.7 \um. North is up and East is left, with a 3x3 \arcsec field of view. The scale is logarithmic in intensity.
          \label{fig:visir}}
      \end{figure}

        During the night of March 26 2005, Q-band imaging was performed using the VLT Imager and Spectrometer for the mid-IR (VISIR, \citealt{2004Msngr.117...12L}). For photometric calibration and PSF determination, the standard stars HD\,50310 and HD\,150798 were observed before and after the science observation respectively. Standard chopping and nodding was employed to remove the atmospheric background emission. The imaging was performed with the Q2 filter ($\lambda_{\rm c}$ = 18.72 \um) in the small field mode (pixel field of view = 0.075$\arcsec$). The observing conditions were fair, airmass $\sim$1.4, and the optical seeing $\sim$0.8$\arcsec$. The achieved sensitivity was 80 before and $\sim$100\,mJy/10$\sigma$1h after the science observation. The full width at half maximum (FWHM) of the point spread function (PSF) was $\sim$0.5\arcsec. The data reduction was performed with a dedicated pipeline, which corrects for various instrumental signatures (see \citealt{2008SPIE.7014E..70P, 2009svlt.conf..261P, Pantin2010}). The reduced image is presented in figure \ref{fig:visir}

        \subsection{Image analysis}\label{sec:visir_obs}
        The VISIR image (fig \ref{fig:visir}) shows a resolved disk that is almost completely spherical, consistent with previous imaging at this wavelength \citep{2003ApJ...598L.111L}. Deviations from spherical symmetry are down to the 10 \% level, which means that inclination must be small. Due to the relatively large size of the PSF at this wavelengths, it is difficult to get a direct constraint on the inclination out of the image. Comparison with a set of inclined, PSF-convolved model images show that it is consistent with an inclination less than $\sim$40 degrees and a position angle along the South-East to North-West.

          Due to the increased sensitivity with respect to previous observations, the image shows resolved emission up to $\sim$1.4\arcsec, corresponding to a physical radius of 145 AU at 103 pc. To characterize this emission, we construct a surface brightness profile. Because the disk is not seen face-on, we construct the surface brightness profile from the image by averaging the surface brightness over an elliptical annulus - rather than a circular one. The shape of the annuli correspond to a disk inclination of 42$^{\circ}$ and a position angle of 145$^{\circ}$. The semi-major axis of this ellipse corresponds to the distance from the star at which the radiation is emitted (radius). Annuli are 0.075\arcsec{} wide, the size of one pixel on the VISIR chip. They are thus much smaller than the width of the VISIR PSF.

      \begin{figure}

        \includegraphics[width=\linewidth]{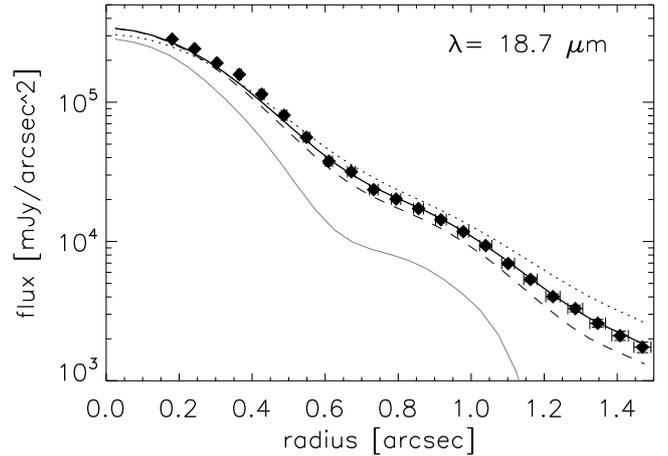}
        \caption[]{Radial emission profiles in the Q band at 18.7 \um, calculated assuming elliptic annuli (see text for details). Displayed are: VISIR image (diamonds) and reference PSF (grey line). The errors on the PSF are displayed as horizontal error bars, those on the noise as vertical ones. Overplotted are PSF-convolved disk models with a varying surface density power law of p=0.0, 1.0 and 2.0 (dotted, solid and dashed respectively).
          NOTE: radius refers to the semi-major axis of the ellipse, which reflects the physical radius where the radiation is emitted.
          \label{fig:radprof}}
      \end{figure}

      Comparing the measured radial intensity profiles to that of the VISIR PSF\footnote{To be consistent, the plotted surface brightness profile of the PSF is also calculated using an elliptical annulus even though it is spherical, and its FWHM in the plot is larger than the 0.5\arcsec previously quoted.} (Fig \ref{fig:radprof}), we see that the disk is clearly resolved. The profile itself is not easily characterized by simple power-laws as is often the case for scattered light images, due to the different emission mechanism (thermal versus scattered light) as well the relatively large size of the PSF.

      Instead, we recognize two components. The profile up to 0.5$\arcsec$ is well described by a Gaussian with a FWHM of 0.73$\arcsec$. After quadratic subtraction of the PSF, we derive a spatial scale of 0.33$\arcsec$ for this central component, corresponding to a physical scale of FWHM = 34 AU at a distance of 103 pc. This number agrees well with the value of 34 $\pm$ 2 found by \cite{2003ApJ...598L.111L} with direct imaging at the same wavelength. This component therefore originates in a region just behind the disk wall, which has a spatial scale of \ong26 AU.

      The second component - outside of 0.5\arcsec - comes from farther out in the disk, and has not been analysed before at this wavelength. Extended emission is seen up to 1.4\arcsec, corresponding to a physical radius of 145 AU at 103 pc. This emission probes the thermal continuum emission from the outer disk surface. The bump around 1.0$\arcsec$ corresponds to the first diffraction ring in the VISIR PSF, and does not represent a real structure in the disk. Because the radial profile traces the disk surface over almost an order of magnitude in radius, we can use it to put limits on the surface density of small grains, which we will do in the next section.

        \subsection{limits on the surface density distribution of small grains}\label{sec:SDP}
        To test if the assumed power law index of p=1 for the surface density profile matches our VISIR image, we compare model images of varying SDP with the observed image, keeping the dust mass and outer radius fixed, as well as the inner disk surface density. Because the pixel size is much smaller than the VISIR PSF, we convolve it with our model images before constructing the radial intensity profiles, and take into account the errors on the width of PSF (FWHM = 0.509$\pm$0.008\arcsec) in the fitting procedure.

        Because we solve for the hydrostatic disk structure, the shape of the radial profile depends only on the surface density distribution (Figure \ref{fig:radprof}). The model is best fit for a surface density power law with index $p=1.1^{+0.4}_{-0.5}$. 

        \section{continuum subtraction}\label{sec:cont_sub}

        \begin{figure}
          \includegraphics[width=\linewidth]{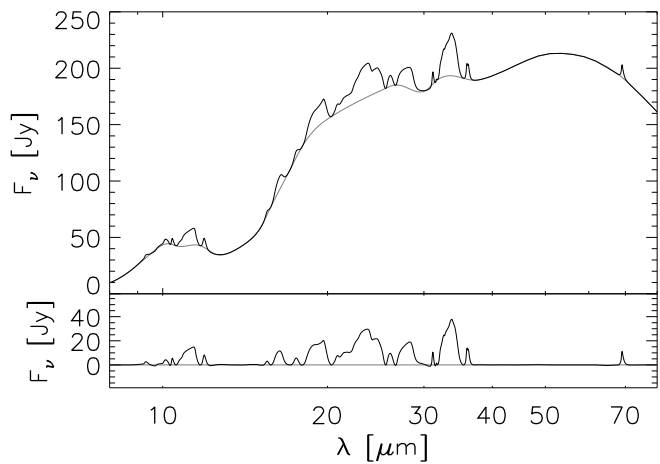}
          \caption[]{Example of continuum subtraction, for model R20. The top panel shows the model spectra (black line) and calculated continuum (grey line). The bottom panel shows the continuum subtracted spectra, with the same color-coding.
            \label{fig:cs_test}}
        \end{figure}

        \begin{figure}
          \includegraphics[width=\linewidth]{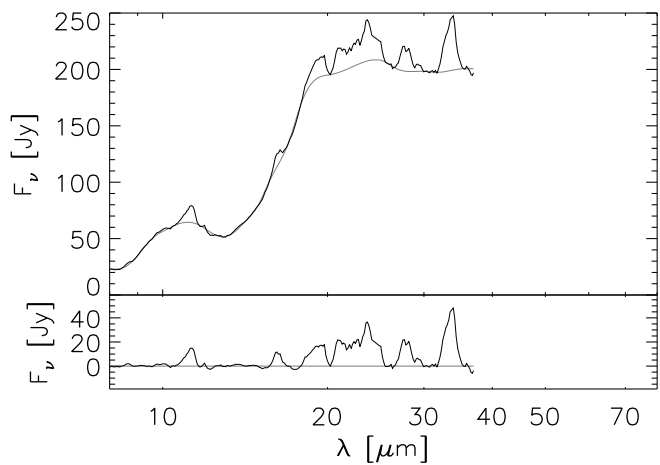}
          \caption[]{Continuum subtraction for the Spitzer spectrum. The top panel shows the model spectra (black line) and calculated continuum (grey line). The bottom panel shows the continuum subtracted spectra, with the same color-coding.
            \label{fig:cs_spitzer}}
        \end{figure}

      To measure feature strengths and compare continuum subtracted spectra, we have to fit and subtract the continuum from both our model spectra and data. An example for model R20 is shown in figure \ref{fig:cs_test}, for the Spitzer spectrum in figure \ref{fig:cs_spitzer}. We are interested in the narrow emission features of forsterite (a crystalline silicate) which are superimposed on much broader features from amorphous silicates, that occupy the same wavelength region - except for the 69 \um feature. The continuum - that includes these broad features - is therefore not easily described by a low-order polynomial.

      The continuum is better fit using a spline function, which is essentially a piecewise polynomial between (arbitrary) continuum points - or knots. At every knot, the polynomials are connected such that they have a continuus derivative. The best result is obtained when knots are placed just outside the features, and the knots ('continuum points') between features are spaced by a distance roughly equal to the feature width.

      The fitted spline follows the shape of the continuum outside the features very well. Inside the features, the continuum is smooth, but has more structure than to be expected from the opacities of amorphous silicates only. Especially the bumps at 12, 27 and 32 \um are artifacts from the continuum subtraction, and not from the underlying amorphous silicate features. They arise because the opacity of forsterite between features is not always zero, especially between the big features and the little 'wings' at 11, 24/28 and 33 \um. These wings are much weaker or not seen in the Spitzer observations (Fig. \ref{fig:cont_sub}), and do appear much weaker in other forsterite opacities \citep{1973PSSBR..55..677S}.

      When comparing the model to the data, we therefore want to exclude these wings, and choose our knots between feature and wings. Although the resulting continuum is not completely smooth - which systematically underestimates the integrated flux - we apply the same continuum subtraction to the observations. This cancels out the systematic error, allowing for a comparison of the shape of the feature, without a systematic error from the wings - which is not in the data. For the 33 \um feature, there is the additional complication that the Spitzer spectrum runs only up to 36.9 \um, and it is not possible to choose a continuum point outside the little wing. We therefore force the tangent at this point to zero to get a reasonable continuum, although this makes the 33 \um flux appear a little stronger than it really is.

      \section{Implementation of temperature-dependent opacities}\label{sec:topac}

      \begin{figure}
        \includegraphics[angle=270,width=\linewidth]{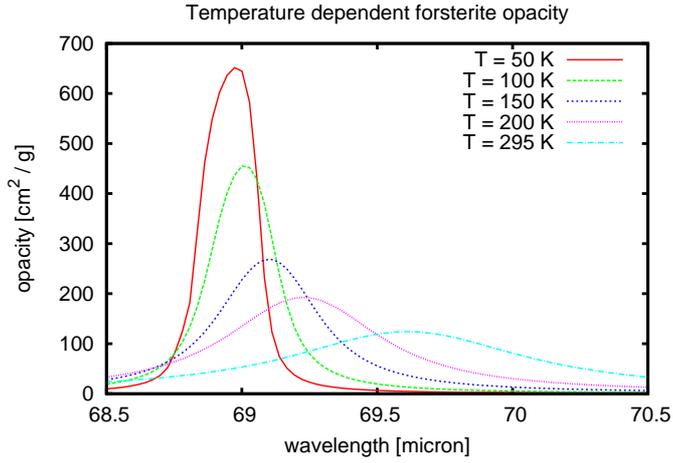}
        \caption[]{Temperature dependence of the forsterite opacity around 69 \um, calculated for the dust grains used in this paper.
          \label{fig:topac}}
      \end{figure}

      When we implement temperature-dependent opacities into the radiative transfer code, we have to take into account that the opacities that set the temperature depend on that same temperature\footnote{Although in the case of forsterite, the changes in the opacities are too small to have a major influence on the disk structure}. We therefore recalculate the opacities at every iteration of the hydrostatic vertical structure loop, to keep the solution selfconsistent.
 
      Optical constants for forsterite have been measured at temperatures of T$_i$= 50, 100, 150, 200 and 295 K \citep{2006MNRAS.370.1599S} for iron free forsterite (Mg$_2$SiO$_4$), but no measurements exist for crystalline olivines with small iron contaminations (Fe$_x$Mg$_{2-x}$SiO$_4$). For temperatures between 50 and 295 K, we calculate the opacity from the optical constants for every temperature component separately, and combine them using a linear interpolation. The opacity at temperature T is calculated as:
      \begin{equation}\label{eq:Topac}
        \kappa(T)=\left( \frac{T_{i+1}-T}{T_{i+1}-T_i}\right)~\kappa_{i} + \left( \frac{T-T_i}{T_{i+1}-T_i} \right)~\kappa_{i+1}
      \end{equation}
      where $\kappa_{i}$ is the opacity at temperature T$_{i}$, and T is between T$_{i}$ and T$_{i+1}$. Below T= 50 K, we use the opacities of 50 K forsterite, and we use the 295 K opacity above a temperature of 295 K. This approach does not take into account the shift in wavelength that might occur outside the 50-295 K range. However the peak shift becomes smaller at lower temperatures \citep{2006MNRAS.370.1599S} and is not so important below 50 K. Temperatures above 295 K are not present in the SED due to the disk gap, except in the inner disk which is does not contribute significantly at 69 \um because it is much hotter.

    \end{appendix}

\end{document}